\documentclass[12pt]{article}

\usepackage{epsfig}
\usepackage{amssymb}
\usepackage{graphicx}
\usepackage{color}
\usepackage{subfigure}
\usepackage{mathtools}
\usepackage[hidelinks]{hyperref}
\usepackage[normalem]{ulem}

\makeatletter
\renewcommand{\theequation}{\thesection.\arabic{equation}}
\@addtoreset{equation}{section} \makeatother
\def\a{\alpha}

\def\d{\delta}
\def\D{\Delta}

\def\ph{\phi}
\def\Ph{\Phi}

\def\l{\lambda}

\def\m{\mu}

\def\th{\theta}

\def\s{\sigma}
\def\S{\Sigma}

\def\O{\Omega}

\def\lt{\left}
\def\rt{\right}

\def\nn{\nonumber}

\DeclareMathOperator{\tr}{Tr}

\def\p{\partial}

\def\la{\langle}
\def\ra{\rangle}

\def\nn{\nonumber}
\def\bul{\noindent$\bullet$}

\def\bea{\begin{eqnarray}}
\def\eea{\end{eqnarray}}

\begin{document}

\thispagestyle{empty}
\begin{flushright}
YGHP-18-06 
\end{flushright}
\vspace{3mm}

\begin{center}
{\Large\bf
Vacua and walls of mass-deformed K\"{a}hler nonlinear sigma models on $Sp(N)/U(N)$} 
\\[12mm]
\vspace{5mm}

  {\large
Masato Arai}$^a$
\footnote{\it  e-mail address: 
arai@sci.kj.yamagata-u.ac.jp
}, 
  {\large  
Anastasia Golubtsova}$^{b \,c}$
\footnote{\it  e-mail address: 
golubtsova@theor.jinr.ru
}, 
  {\large  
Chanyong Park$^d$
}\footnote{\it  e-mail address: 
cyong21@gist.ac.kr
}, \\
{\large  Sunyoung Shin}$^e$
\footnote{\it  e-mail address: 
sihnsy@skku.edu
} 

\vskip 1.5em

{ $^a$\it Faculty of Science, Yamagata University,\\ 
Yamagata 990-8560, Japan   \\
  $^b$\it Bogoliubov Laboratory of Theoretical Physics, JINR, \\
          141980 Dubna, Moscow region, Russia \\
  $^c$\it Dubna State University, Universitetskaya str. 19, Dubna, 141980, Russia\\    
  $^d$\it Department of Physics and Photon Science, 
          Gwangju Institute of Science and Technology, \\
          Cheomdangwagi-ro 123, Buk-gu, Gwangju 61005, 
          Republic of Korea \\
  $^e$\it Institute of Basic Science, Sungkyunkwan University,\\
          Suwon 16419, Republic of Korea   
 }
\vspace{15mm}

{\bf Abstract}\\[5mm]
{\parbox{13cm}{\hspace{5mm}
We study vacua and walls of mass-deformed K\"{a}hler nonlinear sigma models on $Sp(N)/U(N)$. We identify elementary walls with the simple roots of $USp(2N)$ and discuss compressed walls, penetrable walls and multiwalls by using the moduli matrix formalism.
}}
\end{center}
\vfill
\newpage
\setcounter{page}{1}
\setcounter{footnote}{0}
\renewcommand{\thefootnote}{\arabic{footnote}}

\section{Introduction} \label{sec:intro}
\setcounter{equation}{0}

K\"{a}hler and hyper-K\"{a}hler nonlinear sigma models are studied in \cite{DiVecchia:1977nxl,AlvarezGaume:1981hm,Curtright:1979yz}. Massive hyper-K\"{a}hler nonlinear sigma models have a potential which is proportional to the square of a tri-holomorphic Killing vector field of the hyper-K\"{a}hler target space \cite{AlvarezGaume:1983ab}. Fixed points of the Killing vector field are realized as discrete vacua. It was shown that there exist $1/2$ supersymmetric kink solutions that interpolate the discrete vacua \cite{Abraham:1992vb}. A more general potential is possible for a hyper-K\"{a}hler target space of quaternionic dimension two or more, and exact non-singular solutions representing intersecting domain walls are constructed in \cite{Gauntlett:2000bd}. Multi-domain walls are studied in \cite{Gauntlett:2000ib}.

The moduli matrix formalism \cite{Isozumi:2004jc,Isozumi:2004va} was proposed to construct walls systematically in non-Abelian gauge theories with ${\cal N}=2$ supersymmetry in four-dimensional spacetime. The model considered in \cite{Isozumi:2004jc,Isozumi:2004va} becomes massive hyper-K\"{a}hler nonlinear sigma models on the cotangent bundle over the Grassmann manifold $T^\ast G_{N_F,N_C}$\footnote{$G_{N_F,N_C}=\frac{SU(N_F)}{SU(N_C)\times SU(N_F-N_C)\times U(1)}.$} when the gauge coupling is taken to be infinity. In this limit, multiwalls are constructed as well as single walls. Multiwalls are along one spatial direction and their positions depend on moduli parameters and mass parameters. Walls can be compressed to single walls by changing moduli parameters in Abelian gauge theories and in non-Abelian gauge theories. These walls are called compressed walls. A distinguishing feature in the non-Abelian gauge theories is that walls can pass through each other \cite{Isozumi:2004va}. Such walls are called penetrable walls. It was also shown in \cite{Isozumi:2004va} that there is a bundle structure for nondegenerate masses, so that the vacua and the walls are on the K\"{a}hler manifold.

The walls of K\"{a}hler nonlinear sigma models on $SO(2N)/U(N)$ are studied in \cite{Arai:2011gg,Lee:2017kaj}. The Hermitian symmetric space $SO(2N)/U(N)$ is realized as a quadric in the Grassmann manifold $G_{2N,N}$ in accordance with \cite{Hua,Higashijima:1999ki}. As $SO(4)/U(2)\simeq {\bf C}P^1$ and $SO(6)/U(3)\simeq {\bf C}P^3$ \cite{Higashijima:2001vk}, the nonlinear sigma models on $SO(2N)/U(N)$ with $N=2$ and $N=3$ are actually Abelian gauge theories. The walls of the nonlinear sigma models on $SO(2N)/U(N)$ with $N=2,3$ are studied in \cite{Arai:2011gg}.
The walls of the nonlinear sigma models on $SO(2N)/U(N)$ for any $N$ are studied in \cite{Lee:2017kaj}. Penetrable walls, which are related to non-Abelian nature, appear in
$N\geq 4$ cases. The vacua and the walls of $N \leq 7$ cases are presented in pictorial representations where vacua and elementary walls correspond to the vertices and the segments of the representations. It is shown that there is a recurrence of a two dimensional diagram for each $N$ mod 4 in the vacuum structures that are connected to the maximum number of elementary walls. The vacuum structures are proved by induction.

The purpose of this paper is to construct walls of mass-deformed
K\"{a}hler nonlinear sigma models on $Sp(N)/U(N)$\footnote{The result
of this paper is different to the result of \cite{Arai:2011gg}. In
\cite{Arai:2011gg} we did not use the root system of $USp(2N)$ to
analyse the vacua and the walls of the nonlinear sigma models on
$Sp(N)/U(N)$. In this paper we identify the elementary wall
operators with the simple root generators of $USp(2N)$ and find that
the elementary wall operators in \cite{Arai:2011gg} are not correct.
The result of this paper seems to be consistent with the result of
\cite{Eto:2011cv} where kink monopoles are studied in similar models
with $USp(2N)$ global symmetry.}. $Sp(N)\equiv USp(2N)$, or
equivalently $Sp(N)=Sp(N,\mathbf{C})\cap U(2N)$. Unlike $SU(N)$ or
$SO(2N)$, the lengths of the simple roots of $USp(2N)$ are different.
Therefore the operators for the compressed walls of the nonlinear sigma models on $Sp(N)/U(N)$ should be newly
defined. We discuss the definitions of the operators and show that some of multiwalls can be compressed.

Since $Sp(1)/U(1)\simeq {\bf C}P^1 \simeq Q^1$ and $Sp(2)/U(2)\simeq Q^3$ \cite{Higashijima:2001vk}\footnote{${\bf C}P^{N-1}=\frac{SU(N)}{SU(N-1)\times U(1)}$ and $Q^{N-2}=\frac{SO(2N)}{SO(N-2)\times U(1)}$.}, the nonlinear sigma models on $Sp(N)/U(N)$ with $N=1,~2$ are Abelian theories. However, the nonlinear sigma models on $Sp(N)/U(N)$ with $N\geq 3$ are non-Abelian theories, so there exist penetrable walls. We use the pictorial representations proposed in \cite{Lee:2017kaj} to investigate the vacuum structures and the recurrence of two-dimensional diagrams to prove the vacuum structures that are connected to the maximum number of elementary walls by induction.

We follow the convention of \cite{Eto:2011cv,Isaev} for the description of the root systems and corresponding Lie algebras. We also identify the elementary walls with the simple roots of $USp(2N)$
as it is done in \cite{Sakai:2005sp}. In Section \ref{sec:model}, we discuss the nonlinear sigma models on $Sp(N)/U(N)$ and the
moduli matrix formalism. In Section \ref{sec:spn123456}, we study walls of the K\"{a}hler nonlinear sigma models on $Sp(N)/U(N)$ with $N\leq 6$. In Section \ref{sec:vac_max_ele}, we
study the vacuum structures that are connected to the maximum number of elementary walls. In Section
\ref{sec:vac_max_ele_n5}, we make some observations about walls of the nonlinear sigma model on $Sp(5)/U(5)$. In Section
\ref{sec:dissc}, we summarize our results. In Appendix A, we prove the vacuum structures that are connected to the maximum number of elementary walls.

\section{Model}  \label{sec:model}
\setcounter{equation}{0}
The K\"{a}hler nonlinear sigma models on $SO(2N)/U(N)$ and $Sp(N)/U(N)$ can be represented as quadrics in the Grassmann manifold $G_{2N,N}$. The Lagrangian in four dimensions is
written in the ${\cal N}=1$ superfield formalism \cite{Higashijima:1999ki,Higashijima:2001vk,Wess:1992cp}:
\bea
{\mathcal{L}}=\int d^4\th \lt(\Ph_a^{~i}\bar{\Ph}_i^{~b}(e^V)_b^{~a}
-\zeta V_a^{~a} \rt)+\int d^2\th \lt( \Ph_0^{ab}\lt( \Ph_b^{~i}J_{ij}\Ph^{Tj}_{~~a}\rt)+(\mathrm{h.c})\rt), \label{eq:sfld_lag}
\eea
where $\Phi$ is an $N\times 2N$ chiral superfield with the flavor indices $i,j=1,\cdots, 2N$ and the color indices $a,b=1,\cdots N$, $V$ is an $N\times N$ matrix vector superfield in the adjoint representation of $U(N)$ and $\Phi_0^{ab}$ is a chiral superfield under a symmetric representation of $U(N)$.
$\zeta$ is the Fayet-Iliopoulos parameter and we set $\zeta=1$. $J_{ij}$ are invariant tensors defined by
\bea
J=\lt\{
\begin{array}{c}
\s^1 \otimes I_N,\quad SO(2N)/U(N) \\
i\s^2 \otimes I_N, \quad Sp(N)/U(N).
\end{array}
\rt. \label{eq:invtensor}
\eea
The superfields are written in terms of component fields:
\bea
&&\Ph_a^{~i}(y,\th)=\ph_a^{~i}(y)+\sqrt{2}\th\psi_a^{~i}(y)
+\th\th F_a^{~i}(y), \nn\\
&&V_a^{~b}(x,\th,\bar{\th})=2(\th\s^m\bar{\th})A_{ma}^{~~~b}(x)+i(\th\th)(\bar{\th}\bar{\l})_a^{~b}(x)-i(\bar{\th}\bar{\th})(\th\l)_a^{~b}(x)+(\th\th)(\bar{\th}\bar{\th})D_a^{~b}(x), \nn\\
&&\Ph_0^{ab}(y,\th)=\ph_0^{ab}(y)+\sqrt{2}\th\psi_0^{ab}(y)+\th\th F_0^{ab}(y).
\eea
The mass-deformed Lagrangian is obtained by dimensional reduction \cite{Scherk:1978ta}. The bosonic part of the Lagrangian in three dimensions is
\bea
{\mathcal{L}}&=&-|(D_\m\ph)_a^{~i}|^2-|i\ph_a^{~j}M_j^{~i}-i\S_a^{~b}\ph_b^{~i}|^2
+|F_a^{~i}|^2+(D_a^{~b}\ph_b^{~i}\bar{\ph}_i^{~a}-D_a^{~a})\nn\\
&&+\lt((F_0)^{ab}\ph_b^{~i}J_{ij}\ph^{Tj}_{~~a}+\ph_0^{ab}F_b^{~i}J_{ij}\ph^{Tj}_a
+(\ph_0)^{ab}\ph_b^{~i}J_{ij}F^{Tj}_{~~\,a}+\mathrm{(h.c)}  \rt), \label{eq:3dlag}
\eea
where the Greek letter $\m$ is a three-dimensional spacetime index. 
The covariant derivative is defined by
$(D_\m\ph)_a^{~i}=\p_\m\ph_a^{~i}-iA_{\m a}^{~~b}\ph_b^{~i}$. The last term (h.c) stands for the Hermitian conjugate.

The Cartan generators of $SO(2N)$ and $USp(2N)$ are
\bea
H_{I} = e_{I,I} - e_{N+I,N+I}, \quad (I = 1, \cdots N), \label{eq:Ncartan}
\eea
where $e_{I,I}(e_{N+I,N+I})$ is a $2N\times 2N$ matrix whose $(I,I)((N+I,N+I))$ component is one \cite{Eto:2011cv,Isaev}.
The mass matrix can be formulated as
\bea
M=\vec{m}\cdot\vec{H}, \label{eq:mass1}
\eea
with vectors
\bea
&&\vec{m}:=(m_1,m_2\cdots,m_N),\nn\\
&&\vec{H}:=(H_1,H_2,\cdots,H_N). \label{eq:m&H}
\eea
The mass matrix in the basis (\ref{eq:Ncartan}) is
\bea
M=\s_3\otimes\mathrm{diag}(m_1,m_2,\cdots,m_N). \label{eq:mass2}
\eea
Since we are interested in generic mass parameters, we can set $m_1>m_2>\cdots>m_N$ without loss of generality.

Equations of motion for $D$ and $F$ yield the constraints for the Lagrangian (\ref{eq:3dlag})
\bea
&&\ph_a^{~i}\bar{\ph}_i^{~b}-\d_a^{~b}=0, \label{eq:const1}\\
&&\ph_a^{~i}J_{ij}\ph^{Tj}_{~~b}=0, \quad \mbox{(hermitian conjugate)}=0.\label{eq:const2}
\eea
We eliminate the auxiliary fields. The potential term of the model is
\bea
V=|i\ph_a^{~j}M_j^{~i}-i\S_a^{~b}\ph_b^{~i}|^2+4|(\ph_0)^{ab}\ph_b^{~i}|^2.
\eea
The vacuum conditions are
\bea
&&\ph_a^{~j}M_j^{~i}-\S_a^{~b}\ph_b^{~i}=0, \label{eq:vaccond}\\
&&(\ph_0)^{ab}\phi_b^{~i}=0. \label{eq:vaccond2}
\eea
The condition (\ref{eq:vaccond2}) gives $\ph_0=0$ or $\phi=0$. Since the latter solution is
inconsistent with (\ref{eq:const1}), we have $\ph_0=0$.
The scalar field $\S$ can be diagonalized by a $U(N)$ gauge transformation as
\bea
\S=\mathrm{diag}(\S_1,\S_2,\cdots,\S_N). \label{eq:sigma}
\eea
Since $M$ and $\S$ in (\ref{eq:vaccond}) are both diagonal matrices  the vacuum solutions to (\ref{eq:vaccond}) are labelled by
\bea
(\S_1,\S_2,\cdots,\S_N)=(\pm m_1,\pm m_2,\cdots,\pm m_N).\label{eq:vac_label1}
\eea
There exist $2^{N-1}$ vacua in the nonlinear sigma model on $SO(2N)/U(N)$ since the tensor (\ref{eq:invtensor}) is invariant under $O(2N)$ which includes a parity transformation. On the other hand, there exist $2^N$ vacua in the nonlinear sigma model on $Sp(N)/U(N)$. The numbers are the Euler characteristics of the spaces \cite{Witten:1982df}.

To study wall solutions we assume that fields are static and all the fields depend only on the $x_1\equiv x$
coordinate. We also assume that there is Poincar\'{e} invariance on the two-dimensional worldvolume of walls so we can set $A_0=A_2=0$. The energy density along the $x$-direction is
\bea
\mathcal{E}&=&\lt(|(D\ph)_a^{~i}|^2+|\ph_a^{~j}M_j^{~i}-\S_a^{~b}\ph_b^{~i}|^2+4|(\ph_0)^{ab}\ph_b^{~i}|^2\rt) \nn\\
&=&\lt(|(D\ph)_a^{~i}\mp(\ph_a^{~j}M_j^{~i}-\S_a^{~b}\ph_b^{~i})|^2+4|(\ph_0)^{ab}\ph_b^{~i}|^2\rt)\pm {\mathcal{T}} \nn\\
&\geq& \pm {\mathcal{T}}, \label{eq:en-com}
\eea
with $D\equiv D_{\mu=1}$ and
\bea
{\mathcal{T}}=\p(\ph_a^{~i}M_i^{~j}\bar{\ph}_j^{~a}),\quad \partial\equiv \partial_1, \label{eq:tension1}
\eea
which is the tension density of the wall. The tension is
\bea
T=\int^{+\infty}_{-\infty}dx \p\mathrm{Tr}(\ph M \bar{\ph})=\lt[\mathrm{Tr}(\ph M \bar{\ph})\rt]^{+\infty}_{-\infty}.
\eea
The energy density is constrained by (\ref{eq:const1}) and (\ref{eq:const2}).

We choose the upper sign for the BPS equation and the lower sign for the anti-BPS equation in the first squared term in (\ref{eq:en-com}). Then
the BPS equation is
\bea
(D\ph)_a^{~i}-(\ph_a^{~j}M_j^{~i}-\S_a^{~b}\ph_b^{~i})=0. \label{eq:bps}
\eea
We introduce complex matrix functions $S_a^{~b}(x)$ and $f_a^{~i}(x)$, which are defined by
\bea
\S_a^{~b}-iA_a^{~b}=(S^{-1}\p S)_a^{~b},~\ph_a^{~i}=(S^{-1})_a^{~b}f_b^{~i}.
\eea
Then the equation (\ref{eq:bps}) is solved by
\bea
f_b^{~i}=H_{0b}^{~~j}(e^{Mx})_j^{~i}.
\eea
Therefore the solution to the BPS equation (\ref{eq:bps}) is
\bea
\ph_a^{~i}=(S^{-1})_a^{~b}H_{0b}^{~~j}(e^{Mx})_j^{~i}. \label{eq:bps_sol}
\eea
The coefficient matrix $H_0$ is the moduli matrix. $\S$, $A$ and $\ph$ are invariant under the following transformations
\bea
S_a^{\prime~b}=V_a^{~c}S_c^{~b},~~H_{0a}^{\prime~~i}=V_a^{~c}H_{0c}^{~~i}, \label{eq:worldvol}
\eea
where $V\in GL(N,{\mathbf{C}})$. The matrix $V$ defines an equivalent class of $(S,H_0)$. This is named as worldvolume symmetry in the moduli matrix formalism \cite{Isozumi:2004jc,Isozumi:2004va}. (\ref{eq:const1}) and (\ref{eq:const2}) correspond to
\bea
&&H_{0a}^{~~i}(e^{2Mx})_i^{~j}H_{0j}^{\dagger~b}=(SS^\dagger)_a^{~b}\equiv \O_a^{~b},\label{eq:hconst1}\\
&&H_{0a}^{~~i}J_{ij}H^{Tj}_{~~\, b}=0,\quad \mbox{(hermitian conjugate)=0.} \label{eq:hconst2}
\eea
From (\ref{eq:worldvol}) and (\ref{eq:hconst2}) we can learn that moduli matrices $H_0$'s parametrize $Sp(N)/U(N)$.
The tension density (\ref{eq:tension1}) is
\bea
{\mathcal{T}}=\frac{1}{2}\p^2\ln\det\O.
\eea

In the moduli matrix formalism, walls are constructed from elementary walls. The elementary wall operators are the simple root generators of the flavor symmetry. So the elementary walls can be identified with the simple roots \cite{Sakai:2005sp}. We summarize the simple root generators $E_i$, $(i=1,\cdots,N)$ and the simple roots $\vec{\a}_i$ of $SO(2N)$ and $USp(2N)$ following the convention of \cite{Eto:2011cv,Isaev}. The set of vectors $\{\hat{e}_i\}$ is the standard unit vectors $\hat{e}_i\cdot\hat{e}_j=\d_{ij}$:

~~

\bul $SO(2N)$
\bea
&&E_i=e_{i,i+1}-e_{i+N+1,i+N}, \quad (i =1, \cdots, N-1), \nn\\
&&E_N=e_{N-1,2N}-e_{N,2N-1}, \nn\\
&&\vec{\a}_i=\hat{e}_i-\hat{e}_{i+1},  \nn\\
&&\vec{\a}_N=\hat{e}_{N-1}+\hat{e}_N. \label{eq:genrts_so2n}
\eea

\bul $USp(2N)$
\bea
&&E_{i} = e_{i,i+1} - e_{i+N+1,i+N}, \quad (i =1, \dots, N-1), \nn\\
&&E_{N} = e_{N,2N}, \nn\\
&&\vec{\a}_i=\hat{e}_i-\hat{e}_{i+1}, \nn\\
&&\vec{\a}_N=2\hat{e}_N.   \label{eq:genrts_usp2n}
\eea
The Cartan generators (\ref{eq:Ncartan}) and the root generators (\ref{eq:genrts_usp2n}) are related and normalized by
\bea
&&\tr(H_IH_J)=2\d_{IJ},\quad (I,J=1,\cdots,N), \nn\\
&&\tr(H_IE_i)=0,  \nn\\
&&\tr(E_i E_i^\dagger)=\frac{4}{\vec{\a}_i\cdot \vec{\a}_i}.
\eea

In this paper, $\la A \ra$ denotes a vacuum and $\la A \leftarrow B \ra$ denotes a wall which connects vacuum $\la A \ra$ and vacuum $\la B \ra$.

The mass matrix $M$ (\ref{eq:mass1}), which is a linear combination of the Cartan generators, and elementary wall $\la A \leftarrow B \ra$, which is generated by Cartan generator $E_i$ are related by
\bea
c[M,E_i]=c(\vec{m}\cdot\vec{\a}_i)E_i=T_{\la A \leftarrow B\ra}E_i,
\label{eq:mass_ele_pos_step}
\eea
where $c$ is a constant and $T_{\la A \leftarrow B\ra}$ is the tension of wall. The moduli matrix of elementary wall $H_{0\la A \leftarrow B \ra}$, which connects $\la A \ra$ and $\la B \ra$ is
\bea
&&H_{0\la A \leftarrow B \ra}=H_{0\la A \ra}e^{E_i(r)}, \nn\\
&&E_i(r)\equiv e^r E_i,\quad (i=1,\cdots,N), \label{eq:eleop}
\eea
where $E_i$ is an elementary wall operator and $r$ is a complex parameter with
$-\infty < \mathrm{Re}(r) < +\infty$.

Unlike $SU(N)$ and $SO(2N)$, the lengths of the simple roots of $USp(2N)$ are different. Therefore the constant $c$ in (\ref{eq:mass_ele_pos_step}) can be different in some vacuum sectors of the nonlinear sigma models on $Sp(N)/U(N)$.

We first review the formalism for the walls of the nonlinear sigma models on $G_{N_F,N_C}$ and $SO(2N)/U(N)$. In this case $c$ is the same in all the sectors of the vacuum structure. Given the aim of the work \cite{Lee:2017kaj}, it can be fixed as $c=1$ for convenience. Elementary walls can be compressed to single walls. In the nonlinear sigma models on $G_{N_F,N_C}$ and on $SO(2N)/U(N)$, a compressed wall of level $n$ which connects $\la A \ra$ and $\la A^\prime \ra$
is
\bea
&&H_{0\la A\leftarrow A^\prime\ra}=H_{0\la A \ra}e^{[E_{i_1},[E_{i_2},[E_{i_3},[\cdots,[E_{i_n},E_{i_{n+1}}]]]\cdots]](r)},\nn\\
&&(i_m=1,\cdots,N;~m=1,\cdots,n+1). \label{eq:compop}
\eea
A double wall moduli matrix is constructed by multiplying a single wall operator to a single wall moduli matrix. By repeating it, we get a triple wall, a quadruple wall and so on. A multiwall which interpolates $\la A \ra$, $\la A^\prime \ra$,$\cdots$, and $\la B \ra$ is
\bea
&&H_{0\la A \leftarrow A^\prime \leftarrow \cdots \leftarrow B\ra}=H_{0\la A \ra}e^{E_{i_1}(r_1)}e^{E_{i_2}(r_2)}\cdots  e^{E_{i_n}(r_n)},\nn\\
&&(i_m=1,\cdots,N;~m=1,\cdots,n),\label{eq:multiwall}
\eea
where parameters $r_i$ $(i=1,2,\cdots)$ are complex parameters ranging $-\infty < \mathrm{Re}(r_i) < \infty$. Elementary walls pass through each other if
\bea
[E_{i_m},E_{i_n}]=0, \label{eq:penop}
\eea
and these walls are named as penetrable walls \cite{Isozumi:2004va}.

Elementary walls can be identified with simple roots by (\ref{eq:mass_ele_pos_step}) \cite{Sakai:2005sp}. Let root vector $\vec{g}_{\la A_1 \leftarrow A_2 \ra}$ denote the wall which connects vacuum $\la A_1 \ra$ and vacuum $\la A_2 \ra$. The corresponding tension of the wall is  $T_{\la A_1 \leftarrow A_2 \ra}=\vec{m}\cdot \vec{g}_{\la A_1 \leftarrow A_2 \ra}$. Then the elementary wall of (\ref{eq:eleop}) is
\bea
\vec{g}_{\la A \leftarrow B \ra}\equiv c\vec{\a}_i.
\eea
The compressed wall of (\ref{eq:compop}) is
\bea
\vec{g}_{\la A \leftarrow A^\prime \ra}\equiv c\vec{\a}_{i_1}+c\vec{\a}_{i_2}+c\vec{\a}_{i_3}+\cdots c\vec{\a}_{i_n}+c\vec{\a}_{i_{n+1}}.
\eea
The root vectors of the two penetrable elementary walls of (\ref{eq:penop}) are orthogonal
\bea
\vec{\a}_{i_m}\cdot \vec{\a}_{i_n}=0.
\eea

Now we study walls of the nonlinear sigma models on $Sp(N)/U(N)$. In this case, $c=2$ for $i=1,\cdots,N-1$ and $c=1$ for $i=N$ in (\ref{eq:mass_ele_pos_step}). An elementary wall $\la A \leftarrow B^\prime \ra$ is
\bea
\vec{g}_{\la A\leftarrow B^\prime \ra}=c\vec{\a}_i.
\eea
The moduli matrix of $\la A \leftarrow A^{\prime\prime}\ra$, which is a compressed wall of level $n$ is
\bea
&&H_{0\la A\leftarrow A^{\prime\prime}\ra}=H_{0\la A \ra}e^{[E_{i_1},[E_{i_2},[E_{i_3},[\cdots,[E_{i_n},E_{i_{n+1}}]]]\cdots]](r)},\nn\\
&&(i_m=1,\cdots,N-1;~ m=1,\cdots,n+1). \label{eq:compopsp}
\eea
The moduli matrices and the operators are the same as (\ref{eq:compop}) for $i=1,\cdots,N-1$. However, the formula should change for operator $E_N$. As an example, an elementary wall $H_{0\la B \leftarrow B^\prime\ra}=H_{0\la B \ra}e^{E_{N-1}(r)}$ and an elementary wall $H_{0\la B^\prime \leftarrow B^{\prime\prime}\ra}=H_{0\la B^\prime \ra}e^{E_N(r)}$ are compressed to
\bea
H_{0\la B\leftarrow B^{\prime\prime} \ra}=
H_{0\la B \ra}e^{[E_{N-1},[E_{N-1},E_N]](r)},\label{eq:enew1}
\eea
or
\bea
H_{0\la B\leftarrow B^{\prime\prime} \ra}=
H_{0\la B \ra}e^{[[E_N,E_{N-1}],E_{N-1}](r)}.\label{eq:enew2}
\eea
The formulas for multiwalls (\ref{eq:multiwall}) and for penetrable walls (\ref{eq:penop}) hold for walls of nonlinear sigma models on $Sp(N)/U(N)$.

The compressed wall of (\ref{eq:compopsp}) in terms of root vectors is 
\bea \vec{g}_{\la A\leftarrow A^{\prime\prime}\ra}=
2\vec{\a}_{i_1}+2\vec{\a}_{i_2}+2\vec{\a}_{i_3}+\cdots 2\vec{\a}_{i_n}+2\vec{\a}_{i_{n+1}}, 
\eea 
whereas the compressed wall
of (\ref{eq:enew1}) and (\ref{eq:enew2}) is \bea \vec{g}_{\la B\leftarrow B^{\prime\prime} \ra}=2\vec{\a}_{N-1}+\vec{\a}_N.
\eea

In this paper we label the moduli matrices of vacua in descending order as
\bea
&&(\S_1,\S_2,\cdots,\S_{N-1},\S_N)=(m_1,m_2,\cdots,m_{N-1},m_N), \nn\\
&&(\S_1,\S_2,\cdots,\S_{N-1},\S_N)=(m_1,m_2,\cdots,m_{N-1},-m_N), \nn\\
&&(\S_1,\S_2,\cdots,\S_{N-1},\S_N)=(m_1,m_2,\cdots,-m_{N-1},m_N), \nn\\
&&(\S_1,\S_2,\cdots,\S_{N-1},\S_N)=(m_1,m_2,\cdots,-m_{N-1},-m_N), \nn\\
&& \quad \vdots \nn\\
&&(\S_1,\S_2,\cdots,\S_{N-1},\S_N)=(m_1,-m_2,\cdots,-m_{N-1},-m_N), \nn\\
&&(\S_1,\S_2,\cdots,\S_{N-1},\S_N)=(-m_1,m_2,\cdots,m_{N-1},m_N), \nn\\
&&\quad \vdots \nn\\
&&(\S_1,\S_2,\cdots,\S_{N-1},\S_N)=(-m_1,-m_2,\cdots,-m_{N-1},-m_N).
\label{eq:vaclabrule}
\eea

\section{Nonlinear sigma models on $Sp(N)/U(N)$ with $N\leq 6$} \label{sec:spn123456}
\begin{figure}[ht!]
\begin{center}
$\begin{array}{ccc}
\includegraphics[width=3cm,clip]{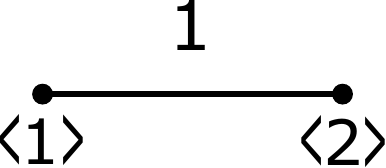}
&~~~~~~&
\includegraphics[width=5cm,clip]{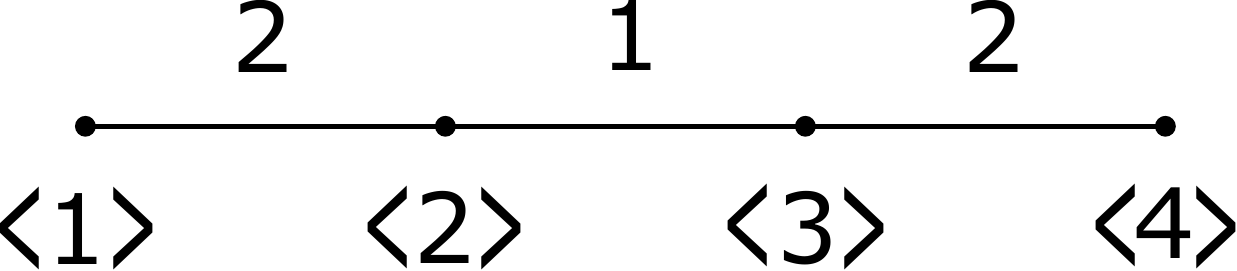}\\
\mathrm{(a)} &~~~~~~& \mathrm{(b)} \\
~~ &  ~~ & ~~ \\
\includegraphics[width=7cm,clip]{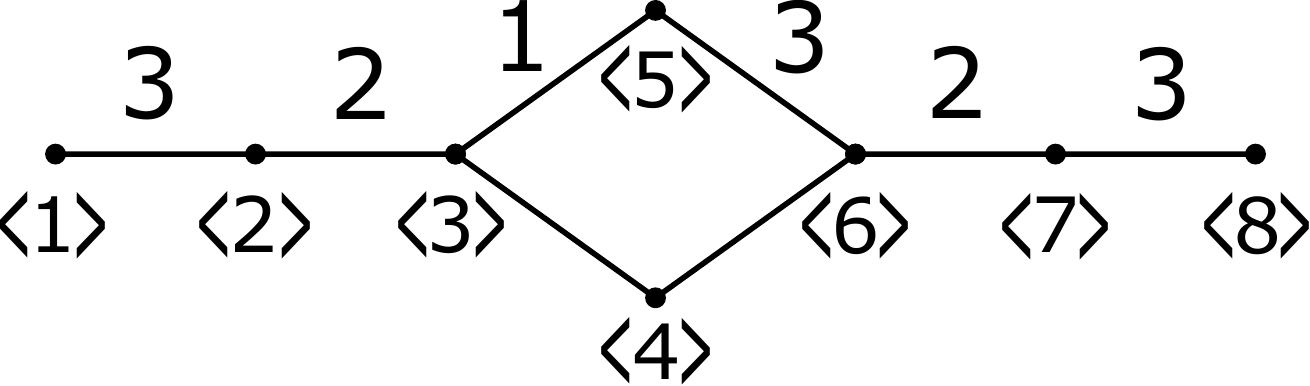}
&~~~~~~&
\includegraphics[width=7cm,clip]{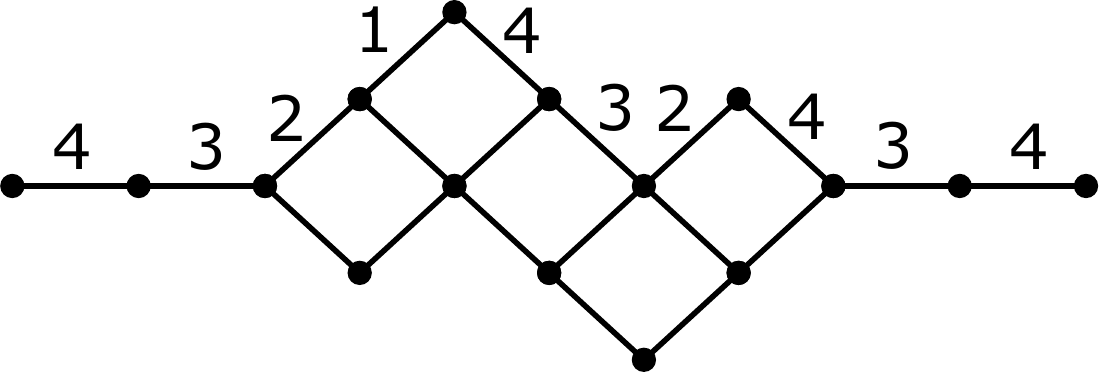}\\
\mathrm{(c)} &~~~~~~& \mathrm{(d)}
\end{array}
$
\end{center}
 \caption{Elementary walls of the nonlinear sigma models on $Sp(N)/U(N)$. (a)$N=1$ (b)$N=2$ (c)$N=3$ and (d)$N=4$. The numbers indicate the subscript $i$'s of roots
  $\vec{\a}_i$. The left-hand side is the limit as
  $x\to +\infty$ and the right-hand side is the limit as $x\to -\infty$. }
 \label{fig:n1n2n3n4}
\end{figure}

\setcounter{equation}{0}
There are two vacua in the nonlinear sigma model on $Sp(1)/U(1)$.
\bea
&&\Ph_{\la 1 \ra}=(1,0),~~\S=m, \nn\\
&&\Ph_{\la 2 \ra}=(0,1),~~\S=-m. \label{eq:vacsol_n1}
\eea
The moduli matrices of the vacua are
\bea
&&H_{0\la 1 \ra}=(1,0),~~\S=m, \nn\\
&&H_{0\la 2 \ra}=(0,1),~~\S=-m. \label{eq:vachn1}
\eea
There is only one wall, which is an elementary wall. The elementary wall operator is
\bea
E_1=e_{1,2},
\eea
and the moduli matrix of the elementary wall is
\bea
H_{0\la 1\leftarrow 2 \ra}=H_{0\la 1\ra}e^{E(r)}=(1,e^r).
\eea
The tension of the wall is
\bea
T_{\la 1 \leftarrow 2 \ra}=\vec{m}\cdot\vec{\a}_1.
\eea
The diagram of the elementary wall is depicted in Figure \ref{fig:n1n2n3n4}(a).

We study walls of the nonlinear sigma model on $Sp(2)/U(2)$. The Cartan generators $H_I$, $(I=1,2)$, the simple root generators $E_i$, $(i=1,2)$, and the simple roots of $USp(4)$ are
\bea
&&H_1=e_{1,1}-e_{3,3},~~
H_2=e_{2,2}-e_{4,4}, \nn\\
&&E_1=e_{1,2}-e_{4,3},~~E_2=e_{2,4}, \nn\\
&&\vec{\a}_1=\hat{e}_1-\hat{e}_2,~~\vec{\a}_2=2\hat{e}_2.
\eea

For $N=2$  the vacuum condition (\ref{eq:vaccond}) gives  rise to 4 vacua, which have the following form
\bea
&&\Ph_{\la 1 \ra}=\lt(
\begin{array}{cccc}
1  &  0  &  0  &  0  \\
0  &  1  &  0  &  0
\end{array}
\rt),~~(\S_1,\S_2)=(m_1,m_2), \nn\\
&&\Ph_{\la 2 \ra}=\lt(
\begin{array}{cccc}
1  &  0  &  0  &  0  \\
0  &  0  &  0  &  1
\end{array}
\rt),~~(\S_1,\S_2)=(m_1,-m_2), \nn\\
&&\Ph_{\la 3 \ra}=\lt(
\begin{array}{cccc}
0  &  0  &  1  &  0  \\
0  &  1  &  0  &  0
\end{array}
\rt),~~(\S_1,\S_2)=(-m_1,m_2), \nn\\
&&\Ph_{\la 4 \ra}=\lt(
\begin{array}{cccc}
0  &  0  &  1  &  0  \\
0  &  0  &  0  &  1
\end{array}
\rt),~~(\S_1,\S_2)=(-m_1,-m_2) . \label{eq:vacsol_n2}
\eea
The moduli matrices of (\ref{eq:vacsol_n2}) are
\bea
&&H_{0\la 1 \ra}=\lt(
\begin{array}{cccc}
1  &  0  &  0  &  0  \\
0  &  1  &  0  &  0
\end{array}
\rt),~~(\S_1,\S_2)=(m_1,m_2), \nn\\
&&H_{0\la 2 \ra}=\lt(
\begin{array}{cccc}
1  &  0  &  0  &  0  \\
0  &  0  &  0  &  1
\end{array}
\rt),~~(\S_1,\S_2)=(m_1,-m_2), \nn\\
&&H_{0\la 3 \ra}=\lt(
\begin{array}{cccc}
0  &  0  &  1  &  0  \\
0  &  1  &  0  &  0
\end{array}
\rt),~~(\S_1,\S_2)=(-m_1,m_2), \nn\\
&&H_{0\la 4 \ra}=\lt(
\begin{array}{cccc}
0  &  0  &  1  &  0  \\
0  &  0  &  0  &  1
\end{array}
\rt),~~(\S_1,\S_2)=(-m_1,-m_2) . \label{eq:vachn2}
\eea

The moduli matrices of elementary walls that connect the vacua (\ref{eq:vachn2}) are
\bea
&&H_{\la 1\leftarrow 2 \ra}=
H_{\la 1 \ra}e^{E_2(r)}=\lt(
\begin{array}{cccc}
1  &  0  &  0  &  0  \\
0  &  1  &  0  &  e^r
\end{array}
\rt),\nn\\
&&
H_{\la 2\leftarrow 3 \ra}=
H_{\la 2 \ra}e^{E_1(r)}=\lt(
\begin{array}{cccc}
1  &  e^r  &  0  &  0  \\
0  &  0  &  -e^r  &  1
\end{array}
\rt),\nn\\
&&
H_{\la 3\leftarrow 4 \ra}=
H_{\la 3 \ra}e^{E_2(r)}=\lt(
\begin{array}{cccc}
0  &  0  &  1  &  0  \\
0  &  1  &  0  &  e^r
\end{array}
\rt). \label{eq:n2ele}
\eea

The wall solution (\ref{eq:bps_sol}) with $H_{\la 1\leftarrow 2 \ra}$ is
\bea
&&\ph_{12}=
\lt(
\begin{array}{cccc}
1   &   0   &   0   &   0   \\
0   &   e^{m_2x}\D^{-1/2}  &  0  &  e^{-m_2x+r}\D^{-1/2}
\end{array}
\rt), \nn\\
&&\D=e^{2m_2x}+e^{-2m_2x+2\mathrm{Re}(r)}. \label{eq:n2eleph12}
\eea
All the phases, which appear due to the $U(1)$ gauge symmetry, are fixed to zero. The wall (\ref{eq:n2eleph12}) has the limits
\bea
&&{x\rightarrow +\infty},\quad \ph_{12}~{\longrightarrow}~\Ph_{\la 1 \ra}, \nn\\
&&{x\rightarrow -\infty},\quad \ph_{12}~{\longrightarrow}~\Ph_{\la 2 \ra},
\eea
as expected. The wall solution (\ref{eq:bps_sol}) with $H_{\la 2\leftarrow 3 \ra}$ is
\bea
&&\ph_{23}=
\lt(
\begin{array}{cccc}
e^{m_1x}\D_1^{-1/2}   &  e^{m_2x+r} \D_1^{-1/2}   &   0   &   0   \\
0   &   0  &  -e^{-m_1x+r}\D_2^{-1/2}  &  e^{-m_2x}\D_2^{-1/2}
\end{array}
\rt), \nn\\
&&\D_1=e^{2m_1x}+e^{2m_2x+2\mathrm{Re}(r)}, \nn\\
&&\D_2=e^{-2m_1x+2\mathrm{Re}(r)}+e^{-2m_2x}\label{eq:n2eleph23}.
\eea
The wall (\ref{eq:n2eleph23}) has the limits
\bea
&&{x\rightarrow +\infty},\quad \ph_{23}~\to~\Ph_{\la 2 \ra}, \nn\\
&&{x\rightarrow -\infty},\quad \ph_{23}~\to~\left(\begin{array}{cccc}
0 & 1 & 0 & 0\\
0 & 0 & 1 & 0
\end{array}
\right).
\eea
Here $\ph_{23}(x\to-\infty)$ is related to vacuum $\Ph_3$ by a $U(N)$ gauge transformation. Therefore $\ph_{23}(x\to-\infty)$ and $\Ph_{3}$ are the same vacuum. We can also see this by making use of worldvolume symmetry. The moduli matrix of $\ph_{23}(x\to-\infty)$ is
\bea
H_{0\la 3 \ra}^\prime=\left(\begin{array}{cccc}
0 & 1 & 0 & 0\\
0 & 0 & 1 & 0
\end{array}
\right),
\eea
which is related to $H_{0\la 3 \ra}$ by worldvolume symmetry
\bea
H_{0\la 3 \ra}^\prime= \lt(
\begin{array}{cc}
0 & 1 \\
1 & 0
\end{array}
\rt)
H_{0\la 3 \ra}.
\eea
Therefore (\ref{eq:n2eleph23}) is the elementary wall which connects vacuum $\la 2 \ra$ and vacuum $\la 3 \ra$. The wall solution (\ref{eq:bps_sol}) with $H_{\la 3\leftarrow 4 \ra}$ is
\bea
&&\ph_{34}=
\lt(
\begin{array}{cccc}
0   & 0   &   1   &   0   \\
0   &   e^{m_2x}\D^{-1/2}  &  0  & e^{-m_2x+r}\D^{-1/2}
\end{array}
\rt), \nn\\
&&\D=e^{2m_2x}+e^{-2m_2x+2\mathrm{Re}(r)}. \label{eq:n2eleph34}
\eea
The wall solution (\ref{eq:n2eleph34}) has the limits
\bea
&&{x\rightarrow +\infty},\quad \ph_{34}~\to~\Ph_{\la 3 \ra}, \nn\\
&&{x\rightarrow -\infty},\quad \ph_{34}~\to~ \Ph_{\la 4 \ra}.
\eea

Tension $T_{\la A \leftarrow B\ra}$ of the wall that connects vacuum $\Ph_{\langle A \rangle}$ and vacuum $\Ph_{\langle B \rangle}$ is obtained from (\ref{eq:vacsol_n2}). The tensions of the elementary walls are
\bea
&&T_{\la 1 \leftarrow 2 \ra}=\vec{m}\cdot\vec{\a}_2,\\
&&T_{\la 2 \leftarrow 3 \ra}=2\vec{m}\cdot\vec{\a}_1,\\
&&T_{\la 3 \leftarrow 4 \ra}=\vec{m}\cdot\vec{\a}_2.
\eea
Therefore the elementary walls are identified with
\bea
&&\vec{g}_{\la 1 \leftarrow 2 \ra}=\vec{g}_{\la 3 \leftarrow 4 \ra}=\vec{\a}_2, \nn\\
&&\vec{g}_{\la 2 \leftarrow 3 \ra}=2\vec{\a}_1.
\eea
The diagram of the elementary walls are depicted in Figure \ref{fig:n1n2n3n4}(b). We omit the coefficients of the simple roots in elementary wall diagrams in this paper.
From the diagram in Figure \ref{fig:n1n2n3n4}(b), one can see how a compressed walls is constructed. From (\ref{eq:compopsp}),
the compressed wall that interpolates $\la 1 \ra$ and $\la 3 \ra$ is
\bea
H_{0\la 1 \leftarrow 3 \ra}=H_{0\la 1 \ra}e^{[[E_2,E_1],E_1](r)}=
\lt(
\begin{array}{cccc}
1   &   0   &   2e^r   &   0   \\
0   &   1   &    0     &   0
\end{array}\rt), \label{eq:comp13}
\eea
and the compressed wall that interpolates $\la 2 \ra$ and $\la 4 \ra$ is
\bea
H_{0\la 2 \leftarrow 4 \ra}=H_{0\la 2 \ra}e^{[E_1,[E_1,E_2]](r)}=
\lt(
\begin{array}{cccc}
1   &   0   &   2e^r   &   0   \\
0   &   0   &    0     &   1
\end{array}\rt). \label{eq:comp24}
\eea
These are the compressed walls of level one.

It can be shown that these compressed walls can be obtained from double walls.
Moduli matrices of double walls $\la 1\leftarrow 2 \leftarrow 3 \ra$ and $\la 2\leftarrow 3 \leftarrow 4 \ra$ are
\bea
&&H_{0\la 1 \leftarrow 2 \leftarrow 3 \ra}=H_{0\la 1 \leftarrow 2 \ra}e^{E_1(r_2)}=
\lt(
\begin{array}{cccc}
1   &  e^{r_2}  &  0   &  0  \\
0   &  1  & -e^{r_1+r_2}  &  e^{r_1}
\end{array}\rt),  \nn\\
&&H_{0\la 2 \leftarrow 3 \leftarrow 4 \ra}=H_{0\la 2 \leftarrow 3 \ra}e^{E_2(r_2)}=
\lt(
\begin{array}{cccc}
1   &  e^{r_1}  &  0   &  e^{r_1+r_2}  \\
0   &  0        & -e^{r_1}  &  1
\end{array}\rt).
\eea

$H_{0\la 1 \leftarrow 2 \leftarrow 3 \ra}$
can be transformed as
\bea
H_{0\la 1 \leftarrow 2 \leftarrow 3 \ra}
&\rightarrow &
\lt(
\begin{array}{cc}
1  &  -e^{r_2} \\
e^{-2r_2} & 1
\end{array}
\rt)\lt(
\begin{array}{cccc}
1   &  e^{r_2}  &  0   &  0  \\
0   &  1  & -e^{r_1+r_2}  &  e^{r_1}
\end{array}\rt) \nn\\
&&=\lt(
\begin{array}{cccc}
1   &  0  &  e^{r+\ln 2}   &  -e^{r-r_2+\ln2}  \\
e^{-2r_2}   &  1+e^{-r_2}  & -e^{r-r_2+\ln 2}  &  e^{r-2r_2+\ln 2}
\end{array}\rt), \label{eq:double123}
\eea
where $r:=r_1+2r_2-\ln2$. The limit of $H_{0\la 1 \leftarrow 2 \leftarrow 3 \ra}$ in (\ref{eq:double123}) as $r_2 \rightarrow +\infty$ with finite $r$  equals to $H_{0\la 1 \leftarrow 3\ra}$ in (\ref{eq:comp13}). Or equivalently,
\bea
H_{0\la 1 \leftarrow 2 \leftarrow 3 \ra}
&=&H_{0\la 1 \ra}e^{E_2(r_1)}e^{E_1(r_2)}\nn\\
&\simeq&H_{0\la 1 \ra}e^{E_2(r_1)}e^{[E_1,E_2](r_1+r_2+i\pi)}e^{[E_1,[E_1,E_2]](r_1+2r_2-\ln 2)} \nn\\
&=&H_{0\la 1 \ra}e^{E_2(r-2r_2+\ln 2)}e^{[E_1,E_2](r-r_2+\ln 2+i\pi)}e^{[E_1,[E_1,E_2]](r)},
\eea
where $r:=r_1+2r_2-\ln2$ and $\simeq$ means the following worldvolume symmetry transformation
\bea
H_{0\la 1 \ra}e^{E_1(r_2)}=
\lt(\begin{array}{cc}
1   &  e^{r_2}  \\
0   &  1
\end{array}\rt)H_{0\la 1 \ra}\simeq H_{0\la 1 \ra}.
\eea
As $r_2\to +\infty$ with finite $r$, $H_{0\la 1 \leftarrow 2 \leftarrow 3 \ra} \to H_{0\la 1 \leftarrow 3 \ra}$.

$H_{0\la 2 \leftarrow 3 \leftarrow 4 \ra}$
transforms as
\bea
H_{0\la 2 \leftarrow 3 \leftarrow 4 \ra}
&\rightarrow &
\lt(
\begin{array}{cc}
1  &  -e^{r_1+r_2} \\
0  & 1
\end{array}
\rt)\lt(
\begin{array}{cccc}
1   &  e^{r_1}  &  0   & e^{r_1+r_2}  \\
0   &  0  & -e^{r_1}  &  1
\end{array}\rt) \nn\\
&&=\lt(
\begin{array}{cccc}
1   &  e^{r_1}  &  e^{r+\ln 2}   &   0  \\
0   &  0  & -e^{r_1}  &  1
\end{array}\rt), \label{eq:double234}
\eea
where $r:=2r_1+r_2-\ln 2$. The limit of $H_{0\la 2 \leftarrow 3 \leftarrow 4 \ra}$ in (\ref{eq:double234}) as $r_1\rightarrow -\infty$ with finite $r$ equals to $H_{0\la 2 \leftarrow 4\ra}$ in (\ref{eq:comp24}). Or equivalently,
\bea
H_{0\la 2 \leftarrow 3 \leftarrow 4 \ra}&=&H_{0\la 2 \ra}e^{E_1(r_1)}e^{E_2(r_2)} \nn\\
&=&H_{0\la 2 \ra}e^{E_2(r_2)}e^{E_1(r_1)}e^{[E_1,E_2](r_1+r_2)}e^{[E_1,[E_1,E_2]](2r_1+r_2-\ln 2+i\pi)} \nn\\
&=&H_{0\la 2 \ra}e^{[E_1,E_2](r_1+r_2)}e^{E_1(r_1)+[E_1,[E_1,E_2]](2r_1+r_2)}e^{[E_1,[E_1,E_2]](2r_1+r_2-\ln 2+i\pi)} \nn\\
&\simeq& H_{0\la 2 \ra} e^{E_1(r_1)+[E_1,[E_1,E_2]](2r_1+r_2)}e^{[E_1,[E_1,E_2]](2r_1+r_2-\ln 2+i\pi)} ,
\eea
where $r:=2r_1+r_2-\ln 2$ and $\simeq$ means the following worldvolume symmetry transformation
\bea
H_{0\la 2 \ra}e^{[E_1,E_2](r_1+r_2)}=
\lt(
\begin{array}{cc}
1   &  e^{r_1+r_2}  \\
0   &  1
\end{array}
\rt)H_{0\la 2 \ra}\simeq H_{0\la 2 \ra}.
\eea
As $r_1\to -\infty$ with finite $r$, $H_{0\la 2 \leftarrow 3 \leftarrow 4 \ra} \to H_{0\la 2 \leftarrow 4 \ra}$.

Triple wall $H_{0\la 1 \leftarrow 2 \leftarrow 3 \leftarrow 4 \ra}$ is
\bea
H_{0\la 1 \leftarrow 2 \leftarrow 3 \leftarrow 4 \ra}
=H_{0\la 1 \leftarrow 2 \leftarrow 3 \ra}e^{E_2(r_3)}
=\lt(
\begin{array}{cccc}
1   &  e^{r_2}  &  0   &   e^{r_2+r_3}  \\
0   &  1  & -e^{r_1+r_2}  &  e^{r_1}+e^{r_3}
\end{array}\rt),
\eea
which consists of three elementary walls $\la 1\leftarrow 2\ra$, $\la 2\leftarrow 3\ra$ and $\la 3\leftarrow 4\ra$. Since $[E_2,[E_1,[E_1,E_2]]]=0$ or equivalently $\vec{\a}_2\cdot(2\vec{\a}_1+\vec{\a}_2)=0$, triple wall $\la 1 \leftarrow 2 \leftarrow 3 \leftarrow 4 \ra$ cannot be compressed to a compressed wall of level two. Instead, elementary wall $\la 1\leftarrow 2\ra$ and compressed wall $\la 2\leftarrow 4\ra$, which is a compressed wall of level one, are penetrable each other or
compressed wall $\la 1\leftarrow 3\ra$, which is a compressed wall of level one, and elementary wall $\la 3\leftarrow 4\ra$ are penetrable each other.

We study walls of the nonlinear sigma model on $Sp(3)/U(3)$. The simple root generators and the simple roots of $USp(6)$ are
\bea
&&E_1=e_{1,2}-e_{5,4},\nn\\
&&E_2=e_{2,3}-e_{6,5}, \nn\\
&&E_3=e_{3,6},
\eea
and
\bea
&&\vec{\a}_1=\hat{e}_1-\hat{e}_2, \nn\\
&&\vec{\a}_2=\hat{e}_2-\hat{e}_3, \nn\\
&&\vec{\a}_3=2\hat{e}_3. \label{eq:n3smplrt}
\eea

The eight vacua of the nonlinear sigma model on $Sp(3)/U(3)$ are labelled in the descending order of (\ref{eq:vaclabrule}):
\bea
&&\la 1 \ra : (\S_1,\S_2, \S_3)=(m_1,m_2, m_3), \nn\\
&&\la 2 \ra : (\S_1,\S_2, \S_3)=(m_1,m_2, -m_3),\nn\\
&&\la 3 \ra : (\S_1,\S_2, \S_3)=(m_1,- m_2, m_3), \nn\\
&&\la 4 \ra : (\S_1,\S_2, \S_3)=(m_1,- m_2,  - m_3),\nn\\
&&\la 5 \ra : (\S_1,\S_2, \S_3)=(-m_1, m_2,  m_3), \nn\\
&&\la 6 \ra : (\S_1,\S_2, \S_3)=(-m_1, m_2,  -m_3), \nn\\
&&\la 7 \ra : (\S_1,\S_2, \S_3)=(-m_1, - m_2,  m_3),\nn\\
&&\la 8 \ra :(\S_1,\S_2, \S_3)=(-m_1, - m_2,  -m_3). \label{eq:vacsol_n3}
\eea
The tensions of elementary walls that connect vacua (\ref{eq:vacsol_n3}) are
\bea
&&T_{\la 1\leftarrow 2\ra}=T_{\la 3\leftarrow 4\ra}=T_{\la 5\leftarrow 6\ra}=T_{\la 7\leftarrow 8\ra}=\vec{m}\cdot\vec{\a}_3, \nn\\
&&T_{\la 2\leftarrow 3\ra}=T_{\la 6\leftarrow 7\ra}=2\vec{m}\cdot\vec{\a}_2,\nn\\
&&T_{\la 3\leftarrow 5\ra}=T_{\la 4\leftarrow 6\ra}=2\vec{m}\cdot\vec{\a}_1.
\eea
Therefore the elementary walls are
\bea
&&\vec{g}_{\la 1 \leftarrow 2\ra}=\vec{g}_{\la 3 \leftarrow 4 \ra}=\vec{g}_{\la 5 \leftarrow 6\ra}=\vec{g}_{\la 7 \leftarrow 8 \ra}=\vec{\a}_3, \nn\\
&&\vec{g}_{\la 2 \leftarrow 3\ra}=\vec{g}_{\la 6 \leftarrow 7 \ra}=2\vec{\a}_2, \nn\\
&&\vec{g}_{\la 3 \leftarrow 5\ra}=\vec{g}_{\la 4 \leftarrow 6 \ra}=2\vec{\a}_1. \label{eq:elert3}
\eea

There are penetrable walls since $\vec{\a}_1\cdot\vec{\a}_3=0$. The diagram of the elementary walls of the nonlinear sigma model on $Sp(3)/U(3)$ are depicted in Figure \ref{fig:n1n2n3n4}(c). In this figure, a pair of penetrable elementary walls makes a parallelogram. A pair of facing sides of the parallelogram are the same simple roots whereas a pair of adjacent sides of the parallelogram are orthogonal simple roots.

We make some observations of walls.
One can guess existence of compressed walls from the wall diagram in Figure \ref{fig:n1n2n3n4}(c). Since
$\vec{g}_{\la 1\leftarrow 2 \ra}\cdot \vec{g}_{\la 2 \leftarrow 3 \ra}\neq 0$, elementary wall $\la 1 \leftarrow 2 \ra$ and elementary wall $\la 2 \leftarrow 3 \ra$ are compressed to compressed wall $\la 1 \leftarrow 3 \ra$, which is a compressed wall of level one. The moduli matrix of $\la 1 \leftarrow 3 \ra$ is
\bea
H_{0\la 1\leftarrow 3 \ra}=H_{0\la 1 \ra}e^{[[E_3,E_2],E_2](r)}. \label{eq:wa13n3}
\eea
One can also see that $\vec{g}_{\la 2\leftarrow 3 \ra}\cdot \vec{g}_{\la 3 \leftarrow 5 \ra}\neq 0$. Therefore  elementary wall $\la 2 \leftarrow 3\ra$ and elementary wall $\la 3\leftarrow 5\ra$ are compressed to compressed wall $\la 2 \leftarrow 5\ra$, which is a compressed wall of level one. The moduli matrix of compressed wall $\la 2\leftarrow 5 \ra$ is
\bea
H_{0\la 2\leftarrow 5 \ra}=H_{0\la 2 \ra}e^{[E_2,E_1](r)}.\label{eq:wa25n3}
\eea

\begin{figure}[h!]
\vspace{2cm}
\begin{center}
$\begin{array}{ccc}
\includegraphics[width=5cm,clip]{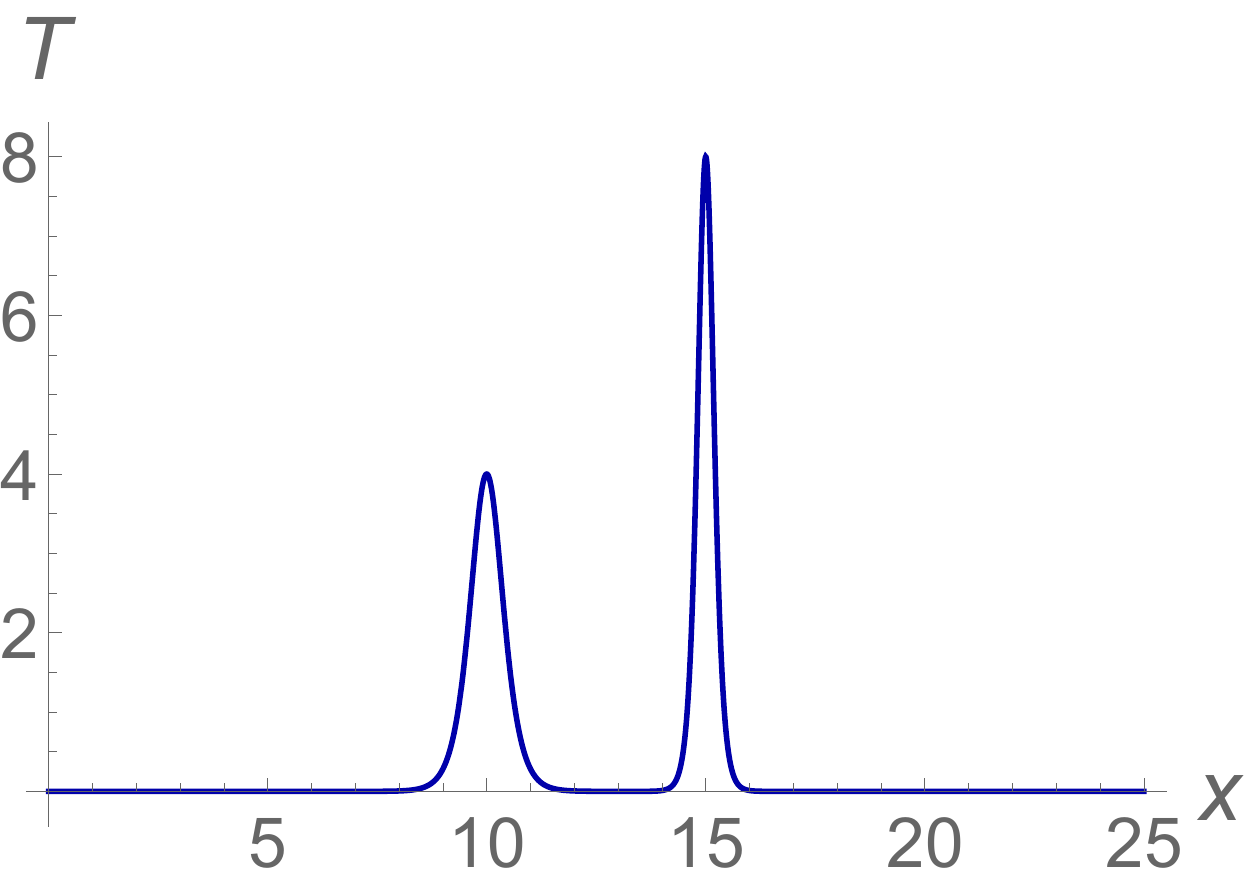}
&
\includegraphics[width=5cm,clip]{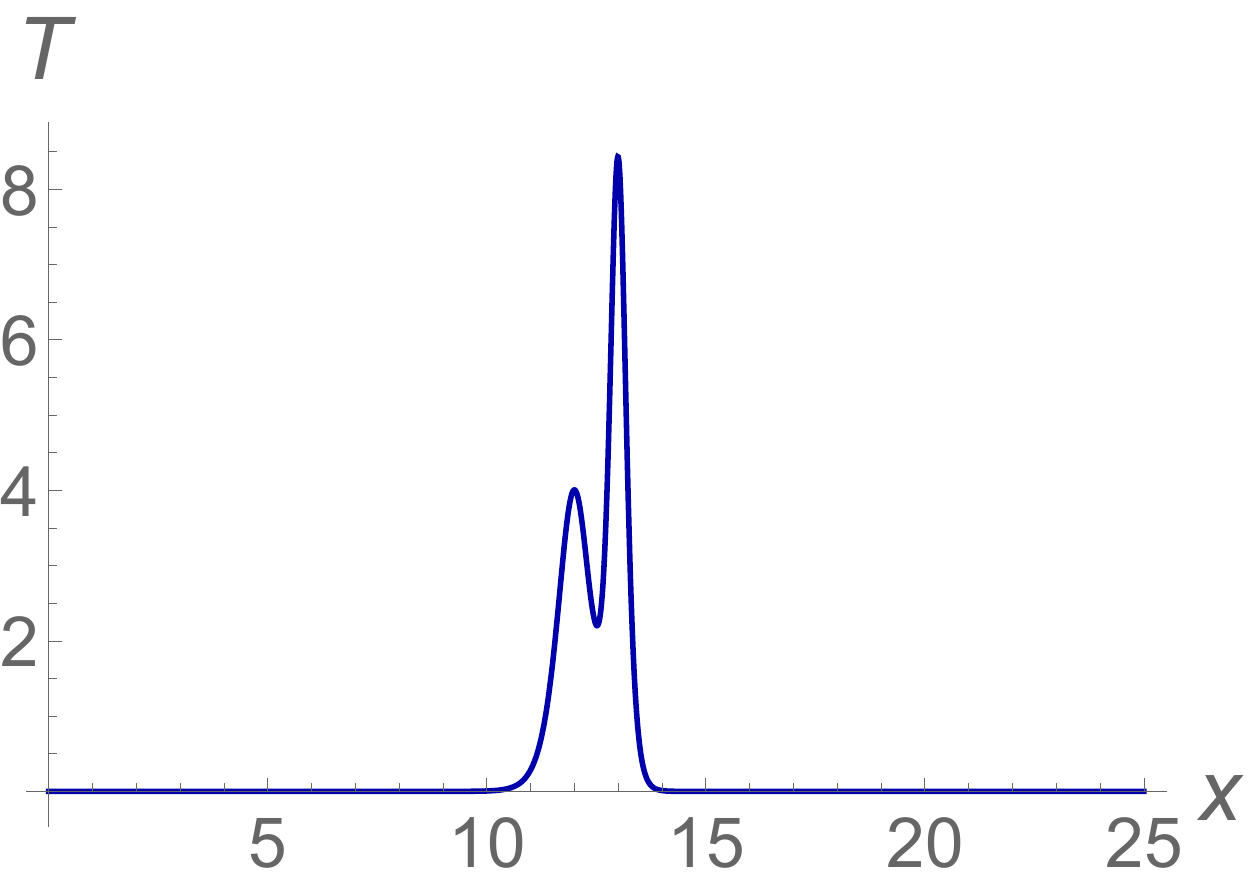}
&
\includegraphics[width=5cm,clip]{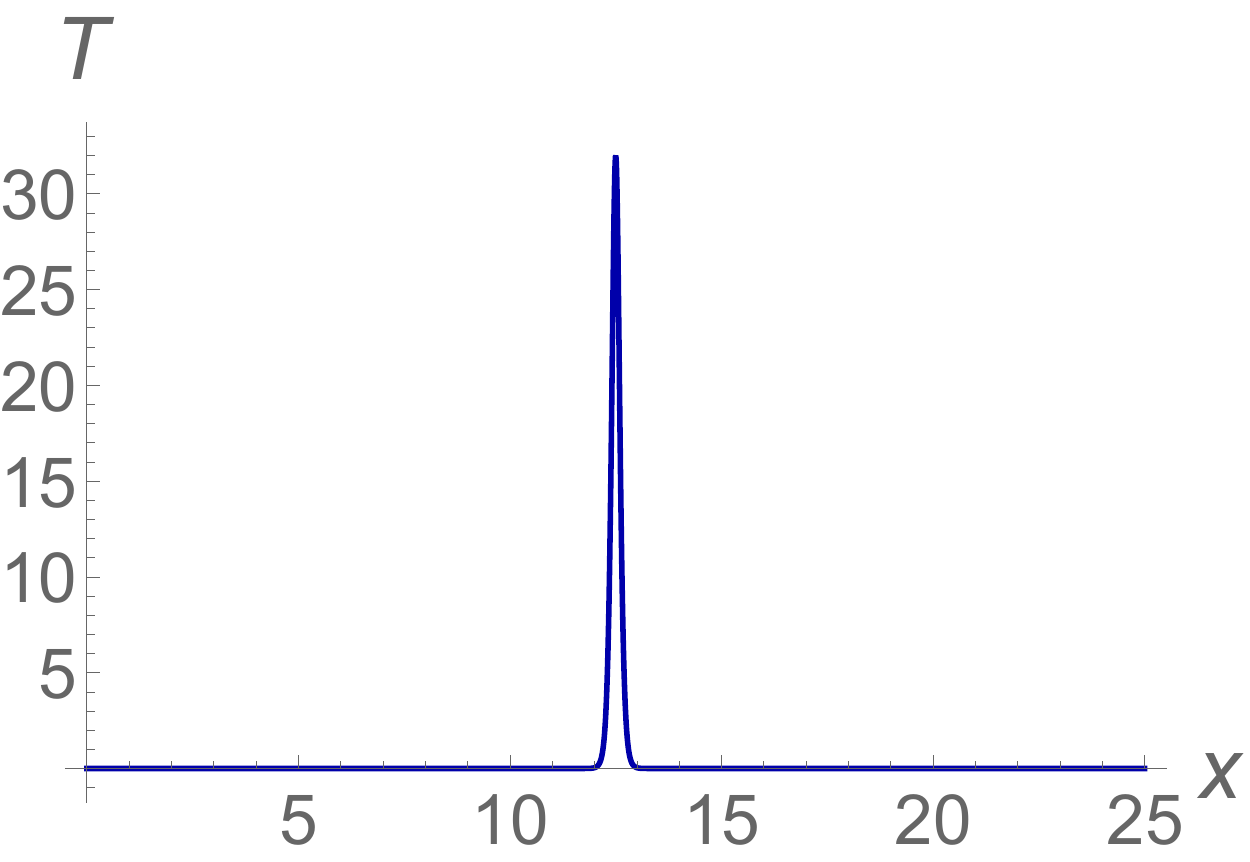}\\
\mathrm{(a)} & \mathrm{(b)} & \mathrm{(c)}
\end{array}
$
\end{center}
 \caption{Double wall $\la 1 \leftarrow  2  \leftarrow 3 \ra$ in $Sp(3)/U(3)$, which consists of two elementary walls $\la 1 \leftarrow  2 \ra$ and
 $\la 2  \leftarrow 3 \ra$. They are compressed to $\la 1  \leftarrow 3 \ra$. $m_1= 8 $, $m_2 = 4$, $m_3 = 2$. (a) $r_1 = 60$, $r_2 = 20$, (b) $r_1 = 52$, $r_2 = 24$, (c) $r_1 = 40$, $r_2 = 30$.}
 \label{fig:cw123}
\end{figure}

Let us consider the moduli matrix of double wall $\la 1 \leftarrow 2 \leftarrow 3\ra$ 
\bea
H_{0\la 1\leftarrow 2 \leftarrow 3 \ra}&=&H_{0\la 1 \ra}e^{E_3(r_1)}e^{E_2(r_2)}, \label{eq:wa123n3}
\eea
and the moduli matrix of double wall $\la 2 \leftarrow 3 \leftarrow 5\ra$ 
\bea
H_{0\la 2\leftarrow 3 \leftarrow 5 \ra}&=&H_{0\la 2 \ra}e^{E_2(r_1)}e^{E_1(r_2)}. \label{eq:wa235n3}
\eea
Double wall $\la 1\leftarrow 2 \leftarrow 3 \ra$ in (\ref{eq:wa123n3}) is
\bea
H_{0\la 1\leftarrow 2 \leftarrow 3 \ra}&=&H_{0\la 1 \ra}e^{E_3(r_1)}e^{E_2(r_2)} \nn\\
&=&H_{0\la 1 \ra}e^{E_2(r_2)}e^{E_3(r_1)}e^{[E_2,E_3](r_1+r_2+i\pi)}
e^{[E_2,[E_2,E_3]](r_1+2r_2-\ln 2)} \nn\\
&\simeq& H_{0\la 1 \ra}e^{E_3(r-2r_2+\ln 2)}e^{[E_2,E_3](r-r_2+\ln 2 +i\pi)}e^{[E_2,[E_2,E_3]](r)}, \label{eq:wa123bn3}
\eea
where $r:=r_1+2r_2-\ln 2$ and $\simeq$ means
\bea
H_{0\la 1 \ra}e^{E_2(r_2)}=
\lt(\begin{array}{ccc}
1  &  0  &  0  \\
0  &  1  & e^{r_2} \\
0  &  0  & 1
\end{array}\rt)H_{0\la 1 \ra}\simeq H_{0\la 1 \ra}.
\eea
As $r_2\to +\infty$ with finite $r$, the limit of $H_{0\la 1\leftarrow 2 \leftarrow 3 \ra}$ equals to $H_{0\la 1\leftarrow 3\ra}$. Double wall $\la 1 \leftarrow 2 \leftarrow 3 \ra$ is plotted in Figure \ref{fig:cw123}.

Double wall $\la 2\leftarrow 3 \leftarrow 5 \ra$ (\ref{eq:wa235n3}) is
\bea
H_{0\la 2\leftarrow 3 \leftarrow 5 \ra}&=&H_{0\la 2 \ra}e^{E_2(r_1)}e^{E_1(r_2)} \nn\\
&=&H_{0\la 2 \ra}e^{E_1(r_2)}e^{E_2(r_1)}e^{[E_2,E_1](r_1+r_2)} \nn\\
&\simeq&H_{0\la 2\ra}e^{E_2(r_1)}e^{[E_1,E_2](r)}, \label{eq:wa235bn3}
\eea
where $r:=r_1+r_2+i\pi$ and $\simeq$ means
\bea
H_{0\la 2 \ra}e^{E_1(r_2)}=
\lt(\begin{array}{ccc}
1  &  e^r  &  0  \\
0  &  1    &  0  \\
0  &  0    &  1
\end{array}\rt)H_{0\la 2 \ra}\simeq H_{0\la 2 \ra}.
\eea
As $r_1\to -\infty$ with finite $r$, the limit of $H_{0\la 2\leftarrow 3 \leftarrow 5 \ra}$ equals to $H_{0\la 2\leftarrow 5\ra}$. Double wall $\la 2\leftarrow 3 \leftarrow 5 \ra$ is compressed to compressed wall $\la 2\leftarrow 5\ra$, which is a compressed wall of level one.

\begin{figure}[h!]
\vspace{2cm}
\begin{center}
$\begin{array}{ccc}
\includegraphics[width=5cm,clip]{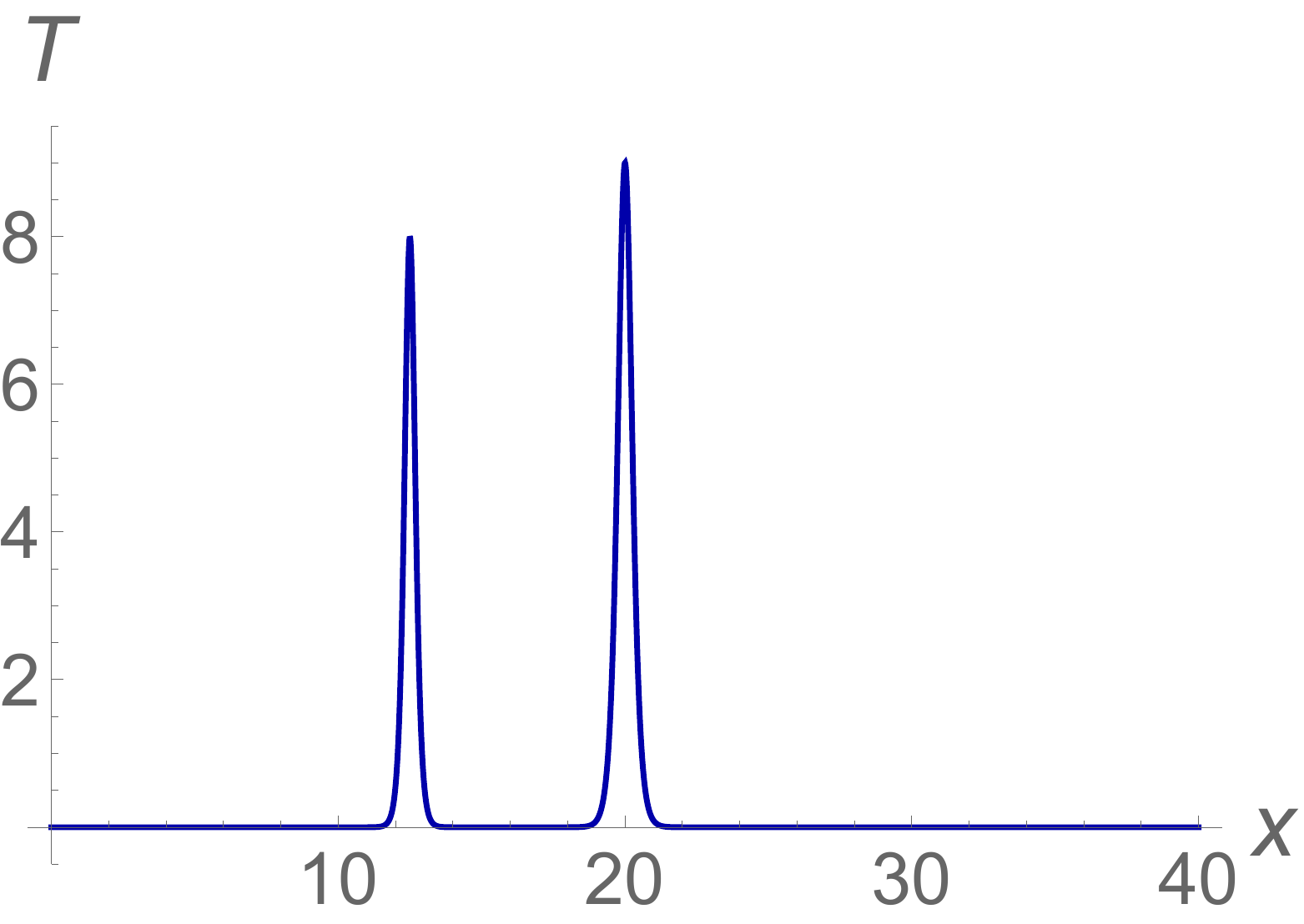}
&
\includegraphics[width=5cm,clip]{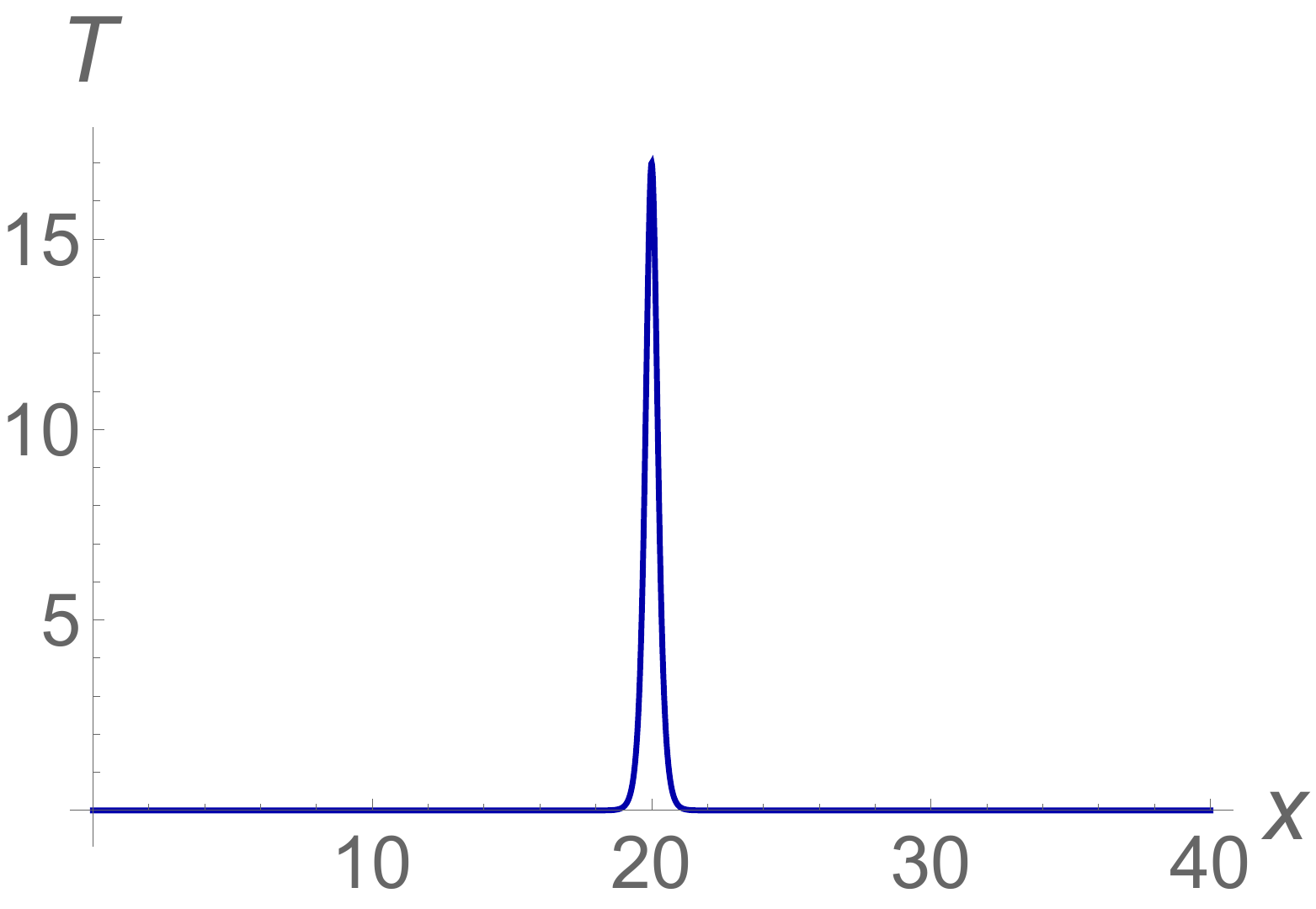}
&
\includegraphics[width=5cm,clip]{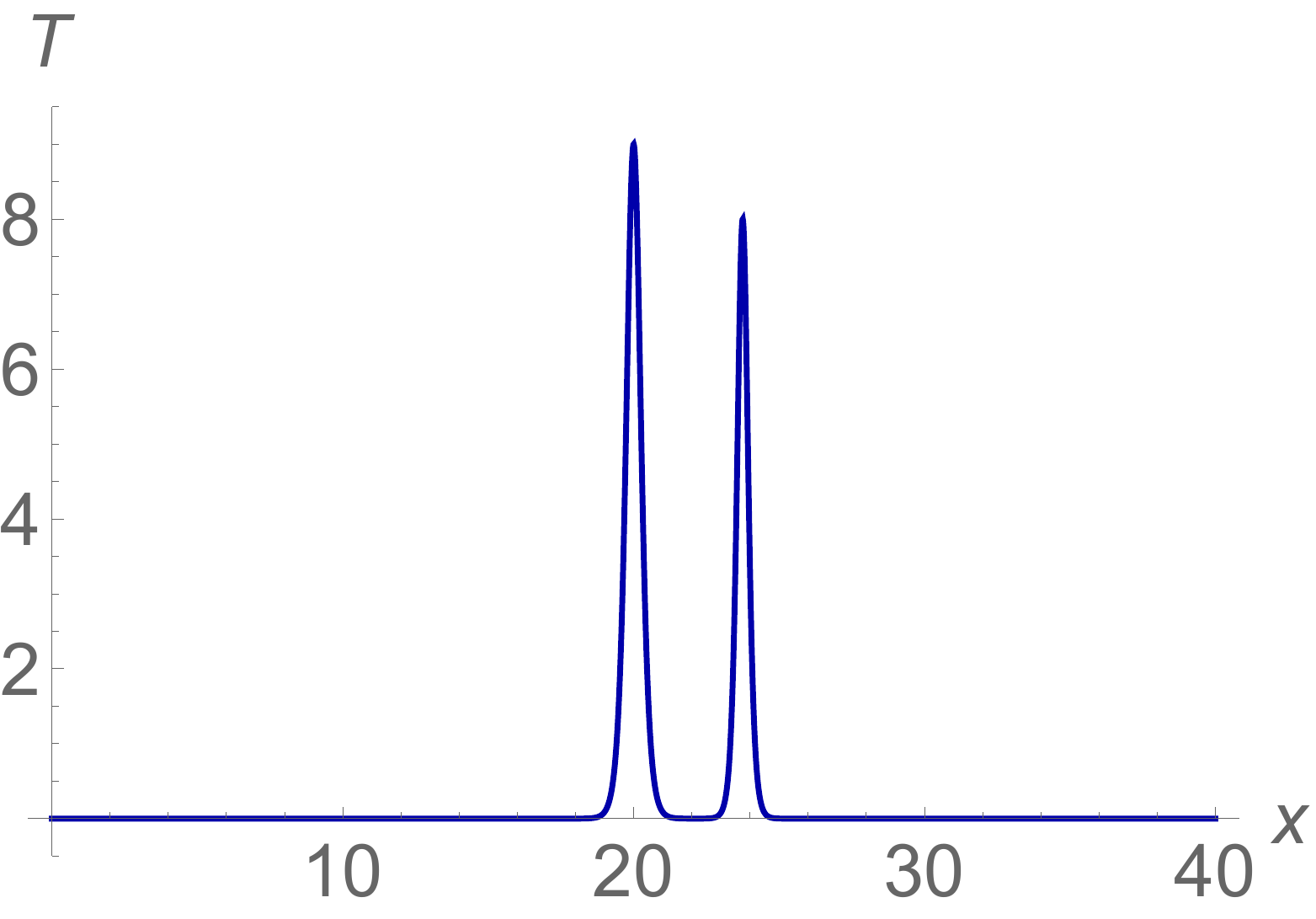}\\
\mathrm{(a)} & \mathrm{(b)} & \mathrm{(c)}
\end{array}
$
\end{center}
 \caption{Double wall $\la 3 \leftarrow  5  \leftarrow 6 \ra$ in $Sp(3)/U(3)$, which consists of two penetrable walls.
 $m_1= 8 $, $m_2 = 5$, $m_3 = 2$. (a) $r_1 = 60$, $r_2 = 50$, (b) $r_1 = 60$, $r_2 = 80$, (c) $r_1 = 60$, $r_2 = 95$.}
 \label{fig:pw356}
\end{figure}

Next we discuss penetrable walls. Since $\vec{g}_{\la 3 \leftarrow 5 \ra}\cdot \vec{g}_{\la 5 \leftarrow 6 \ra}=0$, we can observe elementary wall $\la 3 \leftarrow 5 \ra$ and elementary wall $\la 5 \leftarrow 6 \ra$ pass through each other. Double wall $\la 3 \leftarrow 5 \leftarrow 6 \ra$ is plotted in Figure \ref{fig:pw356}.

The moduli matrix of $\la 1 \leftarrow 5 \ra$, which is a compressed wall of level two is
\bea
H_{0\la 1\leftarrow 5\ra}=H_{0\la 1 \ra}e^{[[[[E_3,E_2],E_2],E_1],E_1](r)}.
\eea
The moduli matrix of triple wall $\la 1\leftarrow 2 \leftarrow 3 \leftarrow 5\ra$ is
\bea
H_{0\la 1\leftarrow 5 \ra}=H_{0\la 1 \ra}e^{E_3(r_1)}e^{E_2(r_2)}e^{E_1(r_3)}.
\eea

We shall consider higher $N$.
Elementary walls can be identified with the simple roots of $USp(2N)$ with proper coefficients. All the compressed single walls and multiwalls can be constructed from the elementary wall configuration. The elementary wall configuration for $N=4$ is
\bea
&&\vec{g}_{\la 5 \leftarrow 9 \ra}=\vec{g}_{\la 6 \leftarrow 10 \ra}=\vec{g}_{\la 7 \leftarrow 11 \ra}=\vec{g}_{\la 8 \leftarrow 12 \ra}=2\vec{\a}_1, \nn\\
\quad\nn\\
&&\vec{g}_{\la 3 \leftarrow 5 \ra}=\vec{g}_{\la 4 \leftarrow 6 \ra}=\vec{g}_{\la 11 \leftarrow 13 \ra}=\vec{g}_{\la 12 \leftarrow 14 \ra}=2\vec{\a}_2, \nn\\\quad\nn\\
&&\vec{g}_{\la 2 \leftarrow 3 \ra}=\vec{g}_{\la 6 \leftarrow 7 \ra}=\vec{g}_{\la 10 \leftarrow 11 \ra}=\vec{g}_{\la 14 \leftarrow 15 \ra}=2\vec{\a}_3, \nn\\\quad\nn\\
&&\vec{g}_{\la 1 \leftarrow 2 \ra}=\vec{g}_{\la 3 \leftarrow 4 \ra}=\vec{g}_{\la 5 \leftarrow 6 \ra}=\vec{g}_{\la 7 \leftarrow 8 \ra}\nn\\
&&=\vec{g}_{\la 9 \leftarrow 10 \ra}=\vec{g}_{\la 11 \leftarrow 12 \ra}=\vec{g}_{\la 13 \leftarrow 14 \ra}=\vec{g}_{\la 15 \leftarrow 16 \ra}=\vec{\a}_4. \label{eq:elert4}
\eea
The diagram of the elementary walls are depicted in Figure \ref{fig:n1n2n3n4}(d). We leave vacuum labels out of diagrams from Figure \ref{fig:n1n2n3n4}(d) onwards.

While the elementary wall diagrams are planar for $N\leq 4$, the diagrams are non-planar for $N\geq 5$. The elementary wall configurations for $N=5$ and $N=6$ are as follows:\\
\bul $N=5$
\bea
&&\vec{g}_{\la 9 \leftarrow 17 \ra}=\vec{g}_{\la 10 \leftarrow 18 \ra}=\vec{g}_{\la 11 \leftarrow 19 \ra}=\vec{g}_{\la 12 \leftarrow 20 \ra} \nn\\
&&=\vec{g}_{\la 13 \leftarrow 21 \ra}=\vec{g}_{\la 14 \leftarrow 22 \ra}=\vec{g}_{\la 15 \leftarrow 23 \ra}=\vec{g}_{\la 16 \leftarrow 24 \ra}=2\vec{\a}_1, \nn\\ \quad\nn\\
&&\vec{g}_{\la 5 \leftarrow 9 \ra}=\vec{g}_{\la 6 \leftarrow 10 \ra}=\vec{g}_{\la 7 \leftarrow 11 \ra}=\vec{g}_{\la 8 \leftarrow 12 \ra} \nn\\
&&=\vec{g}_{\la 21 \leftarrow 25 \ra}=\vec{g}_{\la 22 \leftarrow 26 \ra}=\vec{g}_{\la 23 \leftarrow 27 \ra}=\vec{g}_{\la 24 \leftarrow 28 \ra}=2\vec{\a}_2, \nn\\ \quad\nn\\
&&\vec{g}_{\la 3 \leftarrow 5 \ra}=\vec{g}_{\la 4 \leftarrow 6 \ra}=\vec{g}_{\la 11 \leftarrow 13 \ra}=\vec{g}_{\la 12 \leftarrow 14 \ra} \nn\\
&&=\vec{g}_{\la 19 \leftarrow 21 \ra}=\vec{g}_{\la 20 \leftarrow 22 \ra}=\vec{g}_{\la 27 \leftarrow 29 \ra}=\vec{g}_{\la 28 \leftarrow 30 \ra}=2\vec{\a}_3, \nn\\ \quad\nn\\
&&\vec{g}_{\la 2 \leftarrow 3 \ra}=\vec{g}_{\la 6 \leftarrow 7 \ra}=\vec{g}_{\la 10 \leftarrow 11 \ra}=\vec{g}_{\la 14 \leftarrow 15 \ra} \nn\\
&&=\vec{g}_{\la 18 \leftarrow 19 \ra}=\vec{g}_{\la 22 \leftarrow 23 \ra}=\vec{g}_{\la 26 \leftarrow 27 \ra}=\vec{g}_{\la 30 \leftarrow 31 \ra}=2\vec{\a}_4, \nn\\ \quad\nn\\
&&\vec{g}_{\la 1 \leftarrow 2 \ra}=\vec{g}_{\la 3 \leftarrow 4 \ra}=\vec{g}_{\la 5 \leftarrow 6 \ra}=\vec{g}_{\la 7 \leftarrow 8 \ra} \nn\\
&&=\vec{g}_{\la 9 \leftarrow 10 \ra}=\vec{g}_{\la 11 \leftarrow 12 \ra}=\vec{g}_{\la 13 \leftarrow 14 \ra}=\vec{g}_{\la 15 \leftarrow 16 \ra} \nn\\
&&=\vec{g}_{\la 17 \leftarrow 18 \ra}=\vec{g}_{\la 19 \leftarrow 20 \ra}=\vec{g}_{\la 21 \leftarrow 22 \ra}=\vec{g}_{\la 23 \leftarrow 24 \ra} \nn\\
&&=\vec{g}_{\la 25 \leftarrow 26 \ra}=\vec{g}_{\la 27 \leftarrow 28 \ra}=\vec{g}_{\la 29 \leftarrow 30 \ra}=\vec{g}_{\la 31 \leftarrow 32 \ra}=\vec{\a}_5. \label{eq:elert5}
\label{eq:n5a}
\eea

\bul $N=6$
\bea
&&\vec{g}_{\la 17 \leftarrow 33 \ra}=\vec{g}_{\la 18 \leftarrow 34 \ra}=\vec{g}_{\la 19 \leftarrow 35 \ra}=\vec{g}_{\la 20 \leftarrow 36 \ra} \nn\\
&&=\vec{g}_{\la 21 \leftarrow 37 \ra}=\vec{g}_{\la 22 \leftarrow 38 \ra}=\vec{g}_{\la 23 \leftarrow 39 \ra}=\vec{g}_{\la 24 \leftarrow 40 \ra} \nn\\
&&=\vec{g}_{\la 25 \leftarrow 41 \ra}=\vec{g}_{\la 26 \leftarrow 42 \ra}=\vec{g}_{\la 27 \leftarrow 43 \ra}=\vec{g}_{\la 28 \leftarrow 44 \ra} \nn\\
&&=\vec{g}_{\la 29 \leftarrow 45 \ra}=\vec{g}_{\la 30 \leftarrow 46 \ra}=\vec{g}_{\la 31 \leftarrow 47 \ra}=\vec{g}_{\la 32 \leftarrow 48 \ra}=2\vec{\a}_1, \nn
\eea
\bea
&&\vec{g}_{\la 9 \leftarrow 17 \ra}=\vec{g}_{\la 10 \leftarrow 18 \ra}=\vec{g}_{\la 11 \leftarrow 19 \ra}=\vec{g}_{\la 12 \leftarrow 20 \ra} \nn\\
&&=\vec{g}_{\la 13 \leftarrow 21 \ra}=\vec{g}_{\la 14 \leftarrow 22 \ra}=\vec{g}_{\la 15 \leftarrow 23 \ra}=\vec{g}_{\la 16 \leftarrow 24 \ra} \nn\\
&&=\vec{g}_{\la 41 \leftarrow 49 \ra}=\vec{g}_{\la 42 \leftarrow 50 \ra}=\vec{g}_{\la 43 \leftarrow 51 \ra}=\vec{g}_{\la 44 \leftarrow 52 \ra} \nn\\
&&=\vec{g}_{\la 45 \leftarrow 53 \ra}=\vec{g}_{\la 46 \leftarrow 54 \ra}=\vec{g}_{\la 47 \leftarrow 55 \ra}=\vec{g}_{\la 48 \leftarrow 56 \ra}=2\vec{\a}_2, \nn
\eea
\bea
&&\vec{g}_{\la 5 \leftarrow 9 \ra}=\vec{g}_{\la 6 \leftarrow 10 \ra}=\vec{g}_{\la 7 \leftarrow 11 \ra}=\vec{g}_{\la 8 \leftarrow 12 \ra} \nn\\
&&=\vec{g}_{\la 21 \leftarrow 25 \ra}=\vec{g}_{\la 22 \leftarrow 26 \ra}=\vec{g}_{\la 23 \leftarrow 27 \ra}=\vec{g}_{\la 24 \leftarrow 28 \ra} \nn\\
&&=\vec{g}_{\la 37 \leftarrow 41 \ra}=\vec{g}_{\la 38 \leftarrow 42 \ra}=\vec{g}_{\la 39 \leftarrow 43 \ra}=\vec{g}_{\la 40 \leftarrow 44 \ra} \nn\\
&&=\vec{g}_{\la 53 \leftarrow 57 \ra}=\vec{g}_{\la 54 \leftarrow 58 \ra}=\vec{g}_{\la 55 \leftarrow 59 \ra}=\vec{g}_{\la 56 \leftarrow 60 \ra}=2\vec{\a}_3, \nn
\eea
\bea
&&\vec{g}_{\la 3 \leftarrow 5 \ra}=\vec{g}_{\la 4 \leftarrow 6 \ra}=\vec{g}_{\la 11 \leftarrow 13 \ra}=\vec{g}_{\la 12 \leftarrow 14 \ra} \nn\\
&&=\vec{g}_{\la 19 \leftarrow 21 \ra}=\vec{g}_{\la 20 \leftarrow 22 \ra}=\vec{g}_{\la 27 \leftarrow 29 \ra}=\vec{g}_{\la 28 \leftarrow 30 \ra} \nn\\
&&=\vec{g}_{\la 35 \leftarrow 37 \ra}=\vec{g}_{\la 36 \leftarrow 38 \ra}=\vec{g}_{\la 43 \leftarrow 45 \ra}=\vec{g}_{\la 44 \leftarrow 46 \ra} \nn\\
&&=\vec{g}_{\la 51 \leftarrow 53 \ra}=\vec{g}_{\la 52 \leftarrow 54 \ra}=\vec{g}_{\la 59 \leftarrow 61 \ra}=\vec{g}_{\la 60 \leftarrow 62 \ra}=2\vec{\a}_4, \nn
\eea
\bea
&&\vec{g}_{\la 2 \leftarrow 3 \ra}=\vec{g}_{\la 6 \leftarrow 7 \ra}=\vec{g}_{\la 10 \leftarrow 11 \ra}=\vec{g}_{\la 14 \leftarrow 15 \ra} \nn\\
&&=\vec{g}_{\la 18 \leftarrow 19 \ra}=\vec{g}_{\la 22 \leftarrow 23 \ra}=\vec{g}_{\la 26 \leftarrow 27 \ra}=\vec{g}_{\la 30 \leftarrow 31 \ra} \nn\\
&&=\vec{g}_{\la 34 \leftarrow 35 \ra}=\vec{g}_{\la 38 \leftarrow 39 \ra}=\vec{g}_{\la 42 \leftarrow 43 \ra}=\vec{g}_{\la 46 \leftarrow 47 \ra} \nn\\
&&=\vec{g}_{\la 50 \leftarrow 51 \ra}=\vec{g}_{\la 54 \leftarrow 55 \ra}=\vec{g}_{\la 58 \leftarrow 59 \ra}=\vec{g}_{\la 62 \leftarrow 63 \ra}=2\vec{\a}_5, \nn
\eea
\bea
&&\vec{g}_{\la 1 \leftarrow 2 \ra}=\vec{g}_{\la 3 \leftarrow 4 \ra}=\vec{g}_{\la 5 \leftarrow 6 \ra}=\vec{g}_{\la 7 \leftarrow 8 \ra} \nn\\
&&=\vec{g}_{\la 9 \leftarrow 10 \ra}=\vec{g}_{\la 11 \leftarrow 12 \ra}=\vec{g}_{\la 13 \leftarrow 14 \ra}=\vec{g}_{\la 15 \leftarrow 16 \ra} \nn\\
&&=\vec{g}_{\la 17 \leftarrow 18 \ra}=\vec{g}_{\la 19 \leftarrow 20 \ra}=\vec{g}_{\la 21 \leftarrow 22 \ra}=\vec{g}_{\la 23 \leftarrow 24 \ra} \nn\\
&&=\vec{g}_{\la 25 \leftarrow 26 \ra}=\vec{g}_{\la 27 \leftarrow 28 \ra}=\vec{g}_{\la 29 \leftarrow 30 \ra}=\vec{g}_{\la 31 \leftarrow 32 \ra} \nn\\
&&=\vec{g}_{\la 33 \leftarrow 34 \ra}=\vec{g}_{\la 35 \leftarrow 36 \ra}=\vec{g}_{\la 37 \leftarrow 38 \ra}=\vec{g}_{\la 39 \leftarrow 40 \ra} \nn\\
&&=\vec{g}_{\la 41 \leftarrow 42 \ra}=\vec{g}_{\la 43 \leftarrow 44 \ra}=\vec{g}_{\la 45 \leftarrow 46 \ra}=\vec{g}_{\la 47 \leftarrow 48 \ra} \nn\\
&&=\vec{g}_{\la 49 \leftarrow 50 \ra}=\vec{g}_{\la 51 \leftarrow 52 \ra}=\vec{g}_{\la 53 \leftarrow 54 \ra}=\vec{g}_{\la 55 \leftarrow 56 \ra} \nn\\
&&=\vec{g}_{\la 57 \leftarrow 58 \ra}=\vec{g}_{\la 59 \leftarrow 60 \ra}=\vec{g}_{\la 61 \leftarrow 62 \ra}=\vec{g}_{\la 63 \leftarrow 64 \ra}=\vec{\a}_6. \label{eq:elert6}
\eea

\begin{figure}[ht!]
\begin{center}
\includegraphics[width=10cm,clip]{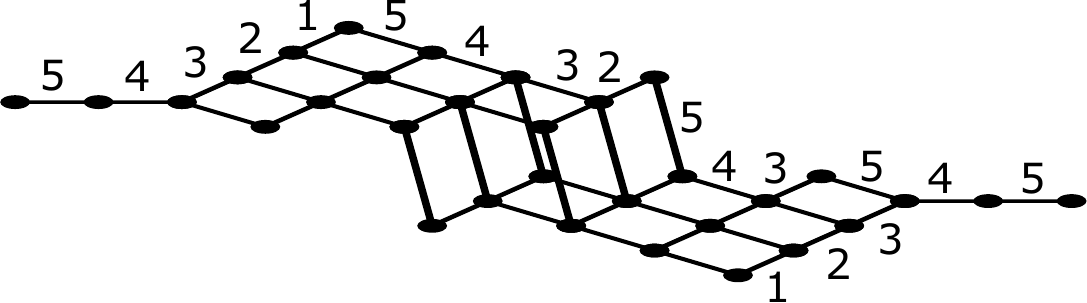}
\end{center}
 \caption{Elementary walls of the nonlinear sigma model on $Sp(5)/U(5)$. The left-hand side is the limit as $x\rightarrow +\infty$ and the right-hand side is the limit as $x\rightarrow -\infty$. }
 \label{fig:n5}
\end{figure}

\begin{figure}[ht!]
\begin{center}
\includegraphics[width=12cm,clip]{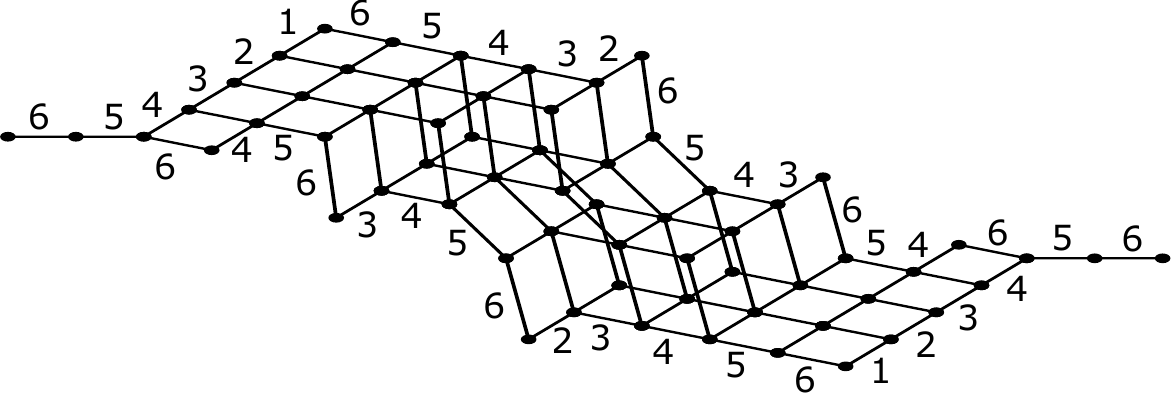}
\end{center}
 \caption{Elementary walls on $Sp(6)/U(6)$. The left-hand side is the limit as $x\to +\infty$ and the right-hand side is the limit as $x\to -\infty$. }
 \label{fig:n6}
\end{figure}

The diagrams of the elementary walls of $N=5$ and $N=6$ cases are depicted in Figure \ref{fig:n5} and Figure \ref{fig:n6}.

\section{Vacua connected to the maximum number of elementary walls} \label{sec:vac_max_ele}
\setcounter{equation}{0}
We study the vacua that are connected to the maximum number of elementary walls. We denote the vacua $\la A \ra$ and $\la B \ra$.  Let $\la A \ra$ be the vacuum near $\la 1 \ra$ and $\la B \ra$ be the vacuum near $\la 2^{N} \ra$. From Figure \ref{fig:n1n2n3n4}, Figure \ref{fig:n5} and Figure \ref{fig:n6}, we make the following observations where $\vec{m}$ denotes simple root $\vec{\alpha}_m$:

\bul $N=1$
\bea
\la 1 \ra \leftarrow \vec{1} \leftarrow \la 2 \ra
\eea

\bul $N=2$
\bea
\vec{2}\leftarrow \la A \ra \leftarrow \vec{1} \leftarrow \la B \ra \leftarrow \vec{2}
\eea

\bul $N=3$
\bea
&&\vec{3} \leftarrow \cdots \leftarrow  \vec{2} \leftarrow \la A\ra \leftarrow \{\vec{1},\vec{3}\} \leftarrow \cdots \nn\\
&&\cdots\leftarrow \{\vec{1},\vec{3} \} \leftarrow \la B \ra \leftarrow \vec{2} \leftarrow \cdots \leftarrow \vec{3}
\eea

\bul $N=4$
\bea
&&\vec{4}\leftarrow \cdots \leftarrow \{\vec{2},\vec{4}\} \leftarrow \la A \ra
\leftarrow \{\vec{1},\vec{3}\}\leftarrow \cdots \nn\\
&&\cdots\leftarrow \{\vec{1},\vec{3}\} \leftarrow \la B \ra \leftarrow \{\vec{2},\vec{4}\} \leftarrow
\cdots \leftarrow \vec{4}
\eea

\bul $N=5$
\bea
&&\vec{5}\leftarrow \cdots \leftarrow \{\vec{2},\vec{4}\} \leftarrow \la A \rangle\leftarrow \{\vec{1},\vec{3},\vec{5}\} \leftarrow \cdots  \nn\\
&&  \cdots \leftarrow \{\vec{1},\vec{3},\vec{5}\} \leftarrow \la B \ra \leftarrow \{\vec{2},\vec{4}\} \leftarrow \cdots \leftarrow \vec{5}
\label{eq:n5b}
\eea

\bul $N=6$
\bea
&&\vec{6}\leftarrow \cdots \leftarrow \{\vec{2},\vec{4},\vec{6}\}\leftarrow \la A \ra \leftarrow \{\vec{1},\vec{3},\vec{5}\} \leftarrow \cdots\nn\\
&&\cdots\leftarrow \{\vec{1},\vec{3},\vec{5}\} \leftarrow \la B \ra \leftarrow \{\vec{2},\vec{4},\vec{6}\}\leftarrow \cdots \leftarrow \vec{6}
\eea

From Figure \ref{fig:n1n2n3n4}, Figure \ref{fig:n5}, Figure \ref{fig:n6}, (\ref{eq:elert3}), (\ref{eq:elert4}), (\ref{eq:elert5}) and (\ref{eq:elert6}), $\la A \ra$ and $\la B \ra$ are identified as follows:\\
\bul $N=3$
\bea
\la A \ra=\la 3 \ra,\quad \la B \ra=\la 6 \ra
\eea
\bul $N=4$
\bea
\la A \ra=\la 6 \ra,\quad \la B \ra=\la 11 \ra
\eea
\bul $N=5$
\bea
\la A \ra=\la 11 \ra,\quad \la B \ra=\la 22 \ra
\eea
\bul $N=6$
\bea
\la A \ra=\la 22 \ra,\quad \la B \ra=\la 43 \ra
\eea
The vacuum labels are not unique since we can change them as we please. Therefore let us label the vacua that are connected to the maximum number of elementary walls as $\la A \ra$ and $\la B \ra$.

The vacuum structures that are connected to the maximum number of elementary walls are as follows:

\bul $N=4m-3 ~~ (m\geq 2)$
\bea
&&\vec{N}\leftarrow \cdots \nn\\
&&\cdots\leftarrow\Big\{\underbrace{\vec{2},\vec{4},\cdots,\overrightarrow{4m-6},\overrightarrow{4m-4}}_{2m-2}\Big\}
\leftarrow \la A \ra \leftarrow \nn\\
&&\leftarrow \Big\{\underbrace{ \vec{1},\vec{3},\cdots,\overrightarrow{4m-5},\overrightarrow{4m-3}}_{2m-1} \Big\} \leftarrow \cdots \nn\\
&&\cdots \leftarrow \Big\{\underbrace{ \vec{1},\vec{3},\cdots,\overrightarrow{4m-5},\overrightarrow{4m-3}}_{2m-1} \Big\} \leftarrow \la B \ra \leftarrow \nn\\
&&\leftarrow \Big\{\underbrace{\vec{2},\vec{4},\cdots,\overrightarrow{4m-6},\overrightarrow{4m-4}}_{2m-2}\Big\} \leftarrow \cdots \nn\\
&&\cdots\leftarrow \vec{N} \label{eq:4m-3}
\eea
\bul $N=4m-2 ~~ (m\geq 2)$
\bea
&&\vec{N}\leftarrow \cdots \nn\\
&&\cdots\leftarrow\Big\{\underbrace{\vec{2},\vec{4},\cdots,\overrightarrow{4m-4},\overrightarrow{4m-2}}_{2m-1}\Big\}
\leftarrow \la A \ra \leftarrow \nn\\
&&\leftarrow \Big\{\underbrace{ \vec{1},\vec{3},\cdots,\overrightarrow{4m-5},\overrightarrow{4m-3}}_{2m-1} \Big\} \leftarrow \cdots \nn\\
&&\cdots \leftarrow \Big\{\underbrace{ \vec{1},\vec{3},\cdots,\overrightarrow{4m-5},\overrightarrow{4m-3}}_{2m-1} \Big\} \leftarrow \la B \ra \leftarrow \nn\\
&&\leftarrow \Big\{\underbrace{\vec{2},\vec{4},\cdots,\overrightarrow{4m-4},\overrightarrow{4m-2}}_{2m-1}\Big\} \leftarrow \cdots \nn\\
&&\cdots\leftarrow \vec{N} \label{eq:4m-2}
\eea
\bul $N=4m-1 ~~ (m\geq 2)$
\bea
&&\vec{N}\leftarrow \cdots \nn\\
&&\cdots\leftarrow\Big\{\underbrace{\vec{2},\vec{4},\cdots,\overrightarrow{4m-4},\overrightarrow{4m-2}}_{2m-1}\Big\}
\leftarrow \la A \ra \leftarrow \nn\\
&&\leftarrow \Big\{\underbrace{ \vec{1},\vec{3},\cdots,\overrightarrow{4m-3},\overrightarrow{4m-1}}_{2m} \Big\} \leftarrow \cdots \nn\\
&&\cdots \leftarrow \Big\{\underbrace{ \vec{1},\vec{3},\cdots,\overrightarrow{4m-3},\overrightarrow{4m-1}}_{2m} \Big\} \leftarrow \la B \ra \leftarrow \nn\\
&&\leftarrow \Big\{\underbrace{\vec{2},\vec{4},\cdots,\overrightarrow{4m-4},\overrightarrow{4m-2}}_{2m-1}\Big\} \leftarrow \cdots \nn\\
&&\cdots\leftarrow \vec{N} \label{eq:4m-1}
\eea
\bul $N=4m ~~ (m\geq 2)$
\bea
&&\vec{N}\leftarrow \cdots \nn\\
&&\cdots\leftarrow\Big\{\underbrace{\vec{2},\vec{4},\cdots,\overrightarrow{4m-2},\overrightarrow{4m}}_{2m}\Big\}
\leftarrow \la A \ra \leftarrow \nn\\
&&\leftarrow \Big\{\underbrace{ \vec{1},\vec{3},\cdots,\overrightarrow{4m-3},\overrightarrow{4m-1}}_{2m} \Big\} \leftarrow \cdots \nn\\
&&\cdots \leftarrow \Big\{\underbrace{ \vec{1},\vec{3},\cdots,\overrightarrow{4m-3},\overrightarrow{4m-1}}_{2m} \Big\} \leftarrow \la B \ra \leftarrow \nn\\
&&\leftarrow \Big\{\underbrace{\vec{2},\vec{4},\cdots,\overrightarrow{4m-2},\overrightarrow{4m}}_{2m}\Big\} \leftarrow \cdots \nn\\
&&\cdots\leftarrow \vec{N} \label{eq:4m}
\eea
(\ref{eq:4m-3}), (\ref{eq:4m-2}), (\ref{eq:4m-1}) and (\ref{eq:4m}) are proved in \ref{sec:app_vac_str}.

\section{Walls of nonlinear sigma model on $Sp(5)/U(5)$} \label{sec:vac_max_ele_n5}
\begin{figure}[ht!]
\vspace{2cm}
\begin{center}
$\begin{array}{ccc}
\includegraphics[width=5cm,clip]{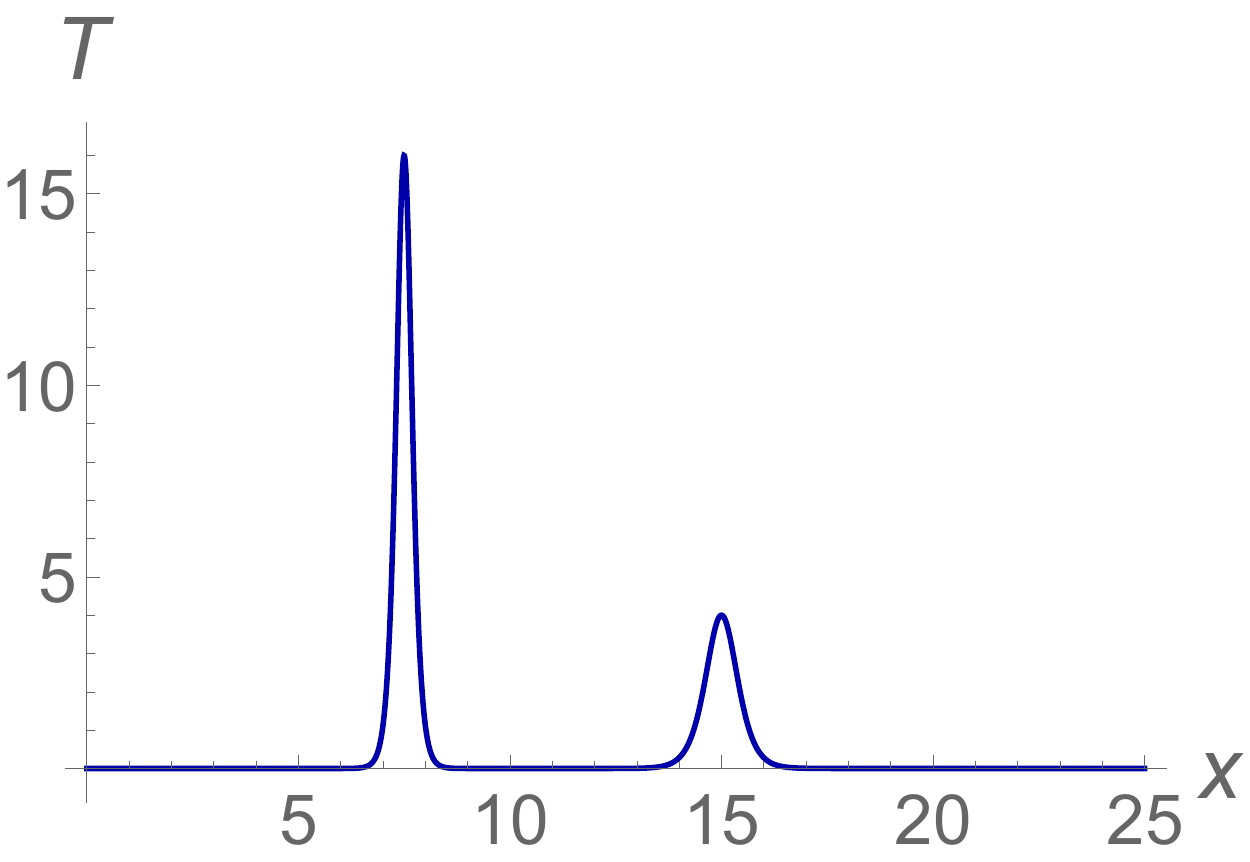}
&
\includegraphics[width=5cm,clip]{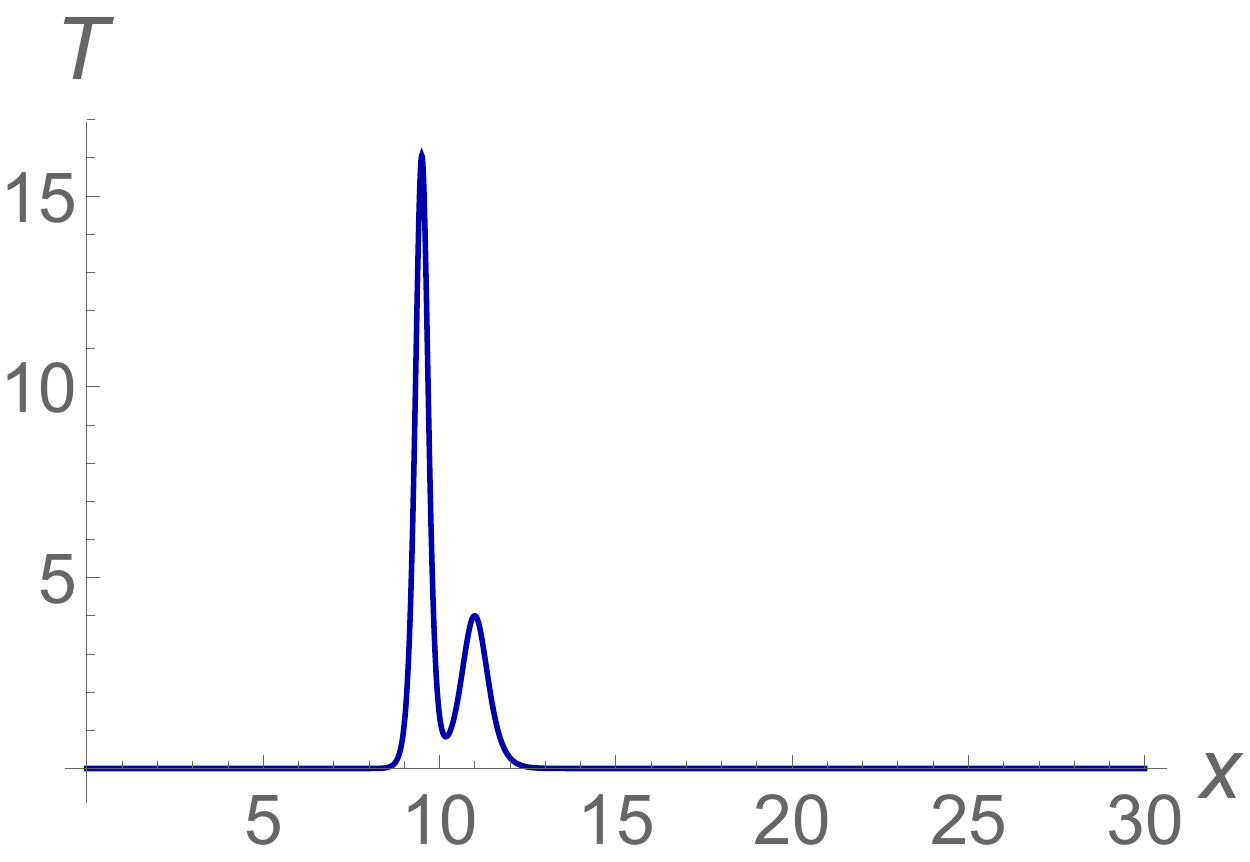}
&
\includegraphics[width=5cm,clip]{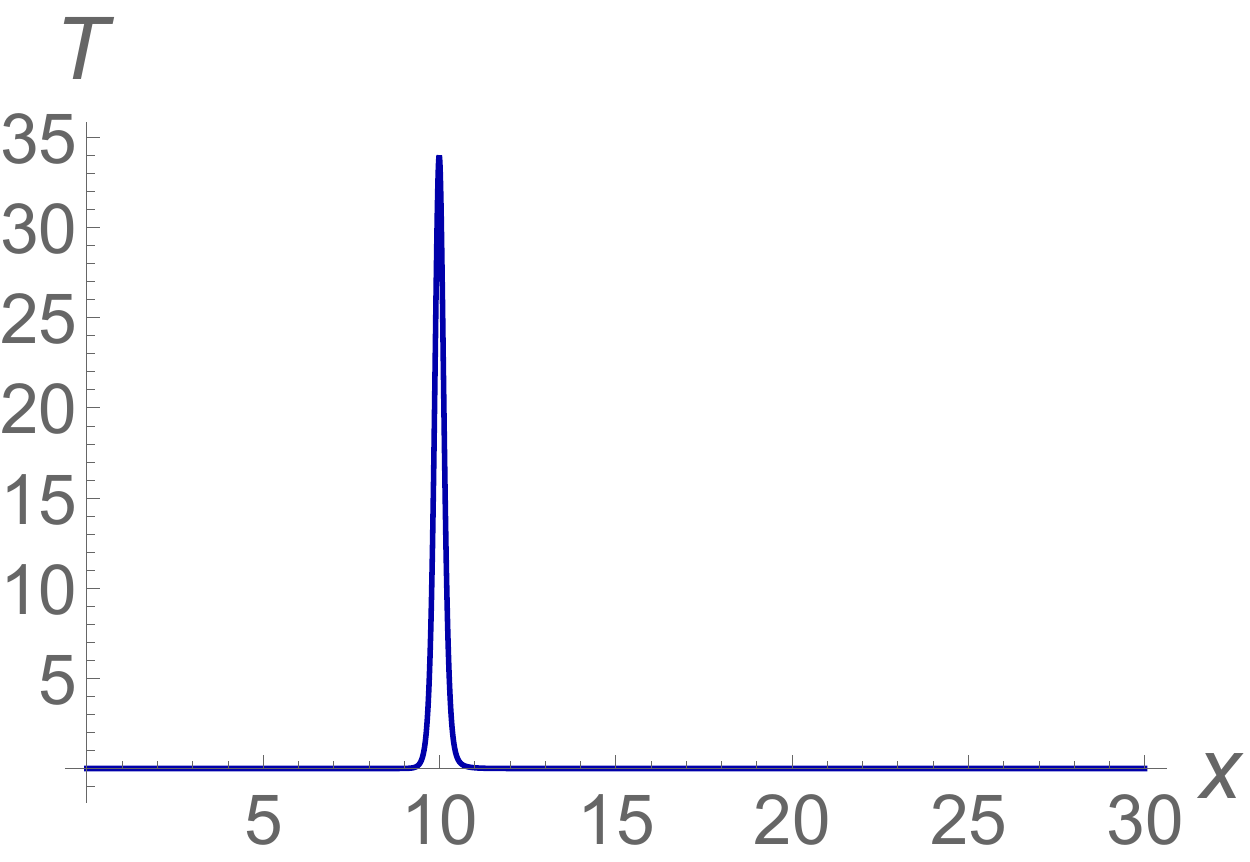}\\
\mathrm{(a)} & \mathrm{(b)} & \mathrm{(c)}
\end{array}
$
\end{center}
 \caption{Double wall $\la 7 \leftarrow  11  \leftarrow 19 \ra$ in $Sp(5)/U(5)$. Elementary walls $\la 7 \leftarrow 11 \ra$ and $\la 11 \leftarrow 19 \ra$ are compressed to $\la 7 \leftarrow 19 \ra$ as $r_1\to -\infty$ with $r_1+r_2=r$(finite).
 $m_1= 12$, $m_2=8$, $m_3=6$, $m_4=4$, $m_5=2$. (a)$r_1=30$, $r_2=30$, (b)$r_1 = 22$, $r_2 = 38$, (c)$r_1=19$, $r_2=41$.}
 \label{fig:sp5_071119}
\end{figure}

\setcounter{equation}{0}
We have studied the vacuum structures that are connected to the maximum number of elementary walls for general $N$. The elementary walls can be compressed or can pass through each other. We discuss some features of elementary walls of nonlinear sigma model on $Sp(5)/U(5)$, which is the simplest nontrivial case. From (\ref{eq:n5a}) and (\ref{eq:n5b}), $\la 11 \ra$ is one of the vacua that are connected to the maximum number of elementary walls. The structure near $\la 11 \ra$ is
\bea
\begin{array}{ccccc}
\begin{array}{ccc}
\la 7 \ra & \leftarrow & \vec{2} \\
\la 10 \ra & \leftarrow & \vec{4}
\end{array}
& \leftarrow & \la 11 \ra & \leftarrow &
\begin{array}{ccc}
 \vec{1} & \leftarrow & \la 19 \ra \\
 \vec{3} & \leftarrow & \la 13 \ra \\
 \vec{5} & \leftarrow & \la 12 \ra
\end{array}
\end{array}
\label{eq:vac11_n5}
\eea
In (\ref{eq:vac11_n5}), $\vec{\a}_2 \cdot \vec{\a}_1\neq 0$. Therefore elementary wall $\la 7 \leftarrow 11 \ra$ and elementary wall $\la 11 \leftarrow 19 \ra$ are compressed to a single wall. Vacuum $\la 7 \ra$ is labelled by $(\S_1,\S_2,\S_3,\S_4,\S_5)=(m_1,m_2,-m_3,-m_4,m_5)$. The moduli matrix for double wall $\la 7 \leftarrow 11 \leftarrow 19 \ra$ is
\bea
H_{0\la 7 \leftarrow 11 \leftarrow 19 \ra}&=&H_{0\la 7 \ra}e^{E_2(r_1)}e^{E_1(r_2)} \nn\\
&=&H_{0\la 7 \ra}e^{E_1(r_2)}e^{E_2(r_1)-[E_1,E_2](r_1+r_2)} \nn\\
&\simeq& H_{0\la 7 \ra}e^{E_2(r_1)-[E_1,E_2](r_1+r_2)}
\eea
where $\simeq$ means
\bea
H_{0\la 7 \ra}e^{E_1(r_2)}=(I_5+e^re_{1,2})H_{0\la 7 \ra}\simeq H_{0\la 7 \ra}.
\eea
As $r_1\to -\infty$ with $r_1+r_2=r$(finite), $H_{0\la 7\leftarrow 11\leftarrow 19 \ra} \to H_{0\la 7\leftarrow 19 \ra}$. Double wall $\la 7\leftarrow 11\leftarrow 19 \ra$ is compressed to compressed wall $\la 7\leftarrow 19 \ra$, which is a compressed wall of level one. This is depicted in Figure \ref{fig:sp5_071119}.

\begin{figure}[ht!]
\vspace{2cm}
\begin{center}
$\begin{array}{ccc}
\includegraphics[width=5cm,clip]{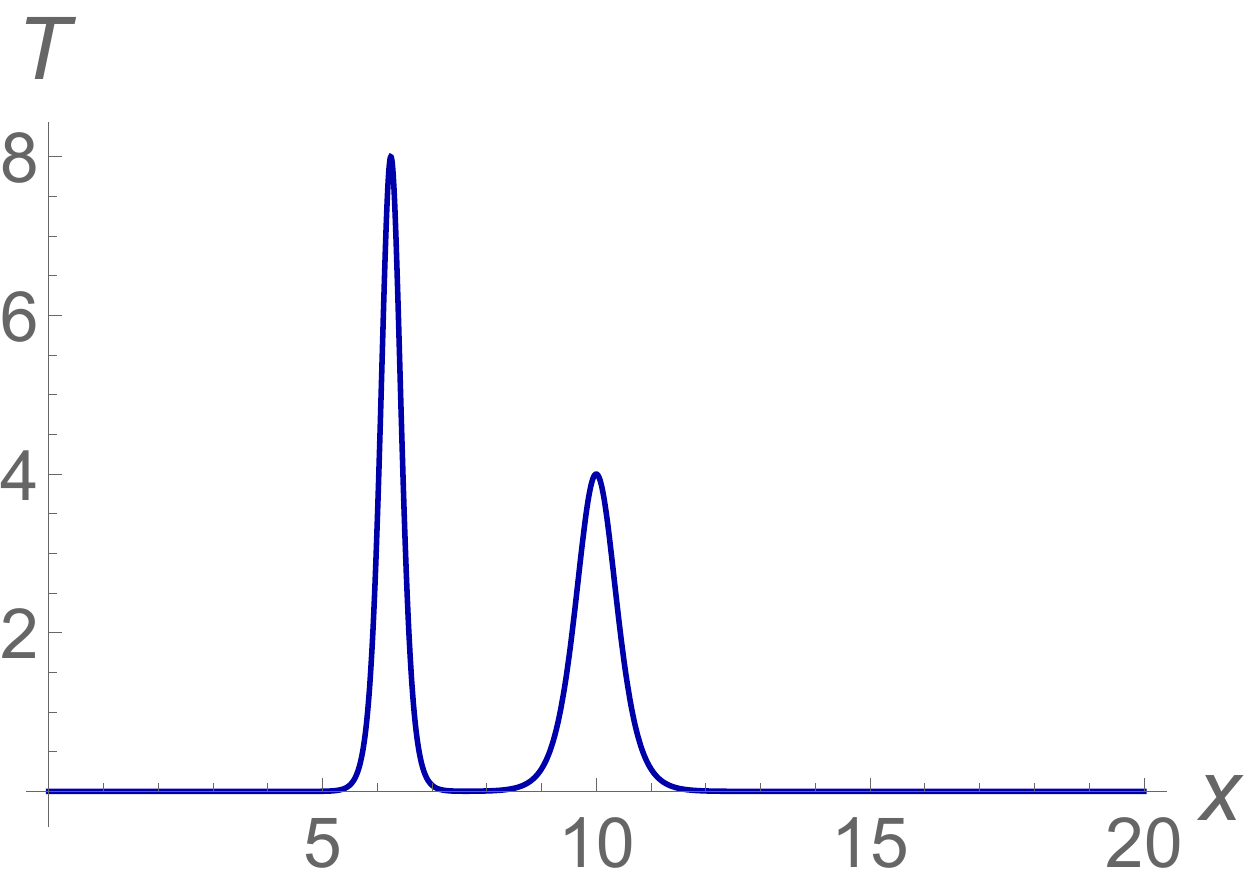}
&
\includegraphics[width=5cm,clip]{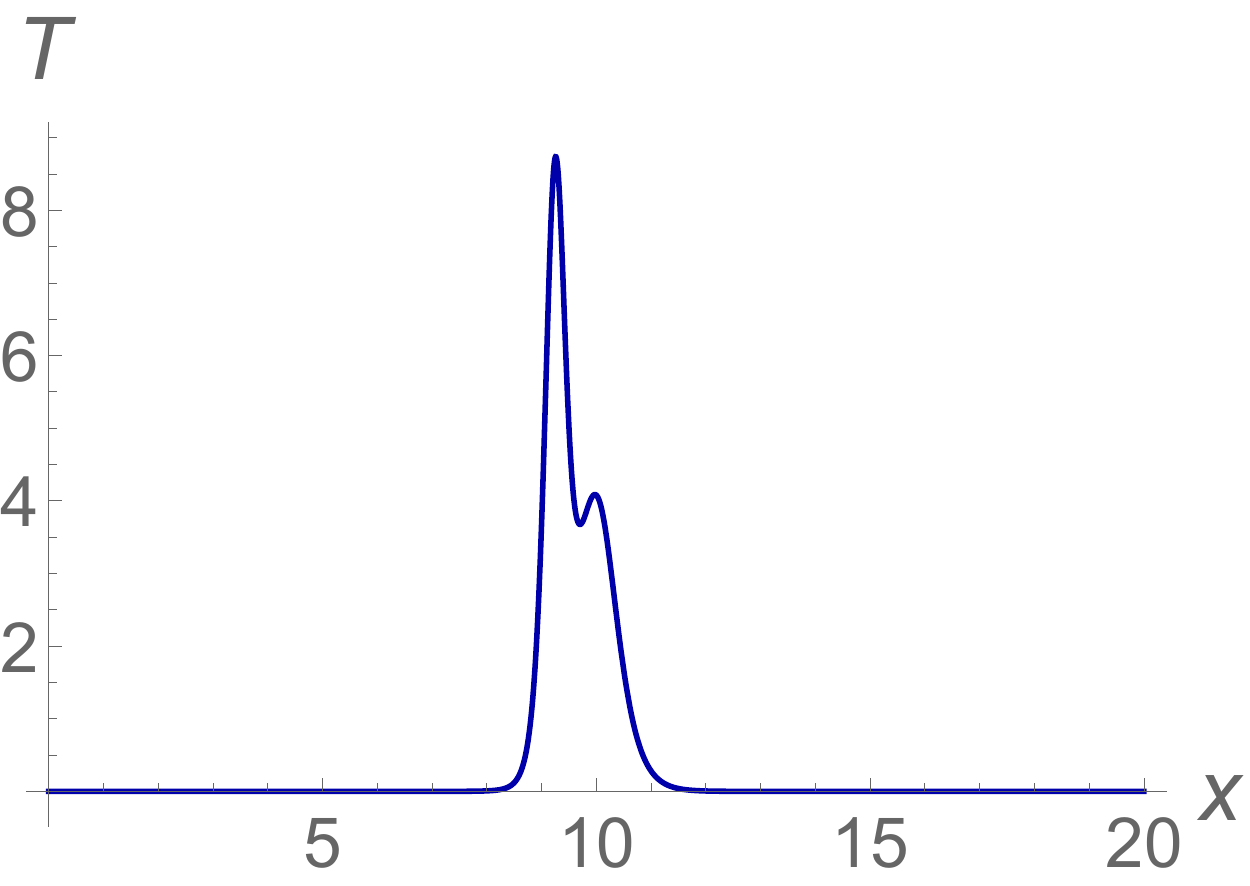}
&
\includegraphics[width=5cm,clip]{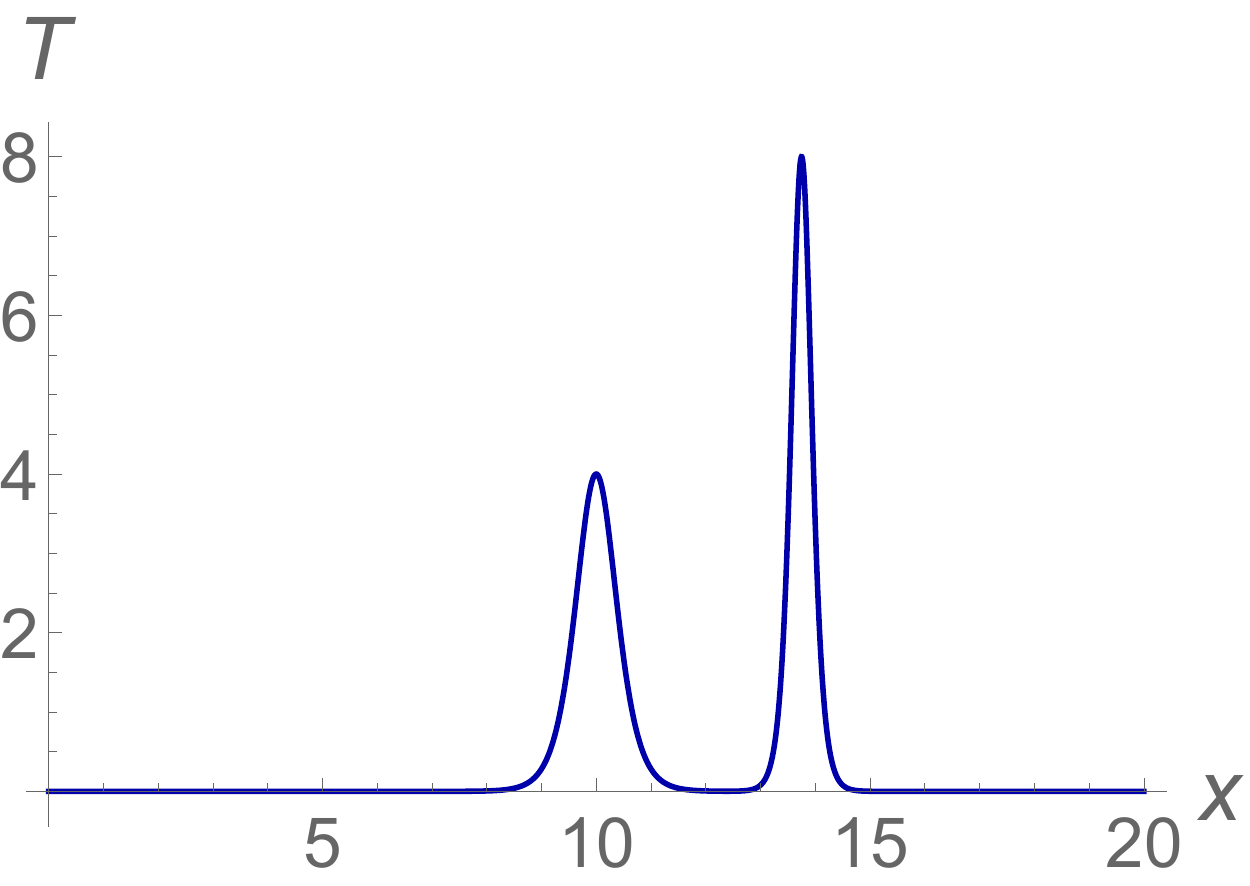}\\
\mathrm{(a)} & \mathrm{(b)} & \mathrm{(c)}
\end{array}
$
\end{center}
 \caption{Double wall $\la 7 \leftarrow  11  \leftarrow 12 \ra$ in $Sp(5)/U(5)$. Elementary walls $\la 7 \leftarrow 11 \ra$ and $\la 11 \leftarrow 12 \ra$ pass through each other.
 $m_1= 12$, $m_2=8$, $m_3=6$, $m_4=4$, $m_5=2$. (a)$r_1=20$, $r_2=25$, (b)$r_1 = 20$, $r_2 = 37$, (c)$r_1=20$, $r_2=55$.}
 \label{fig:sp5_071112}
\end{figure}

In (\ref{eq:vac11_n5}) $\vec{\a}_2 \cdot \vec{\a}_5=0$. Therefore elementary wall $\la 7 \leftarrow 11 \ra$ and elementary wall $\la 11 \leftarrow 12 \ra$ are penetrable. The moduli matrix of double wall $\la 7 \leftarrow 11 \leftarrow 12 \ra$, which consists of two penetrable elementary walls $\la 7 \leftarrow 11 \ra$ and $\la 11 \leftarrow 12 \ra$ is
\bea
&&H_{0\la 7\leftarrow 11 \leftarrow 12 \ra}\nn\\
&&=H_{0\la 7\leftarrow 11 \ra}e^{E_5(r_2)}=H_{0\la 7 \ra}e^{E_2(r_1)}e^{E_5(r_2)}=H_{0\la 7 \ra}e^{E_5(r_2)}e^{E_2(r_1)} \nn\\
&&=H_{0\la 7\leftarrow 8 \ra}e^{E_2(r_1)}.
\eea
This is depicted in Figure \ref{fig:sp5_071112}.

\section{Conclusion}\label{sec:dissc}

We have studied the vacua and the walls of mass-deformed K\"{a}hler nonlinear sigma models on $Sp(N)/U(N)$ by using the moduli matrix formalism. For $N=1$ and $N=2$, the nonlinear sigma models on $Sp(N)/U(N)$ are Abelian theories, where single walls are compressed to compressed walls while penetrable walls are not allowed. On the other hand, for $N\geq 3$, the nonlinear sigma models on $Sp(N)/U(N)$ are non-Abelian theories so there exist penetrable walls, which lead to a unique vacuum configuration for each $N$. We have proved the vacuum structures that are connected to the maximum number of elementary walls by induction.
\\

\noindent {\bf Acknowledgement}

We thank B-H. Lee for early participation. We would like to thank S. Krivonos, A. Sutulin and N. Tyurin for helpful comments. AG would also like to thank the Asian Pacific Center For Theoretical Physics APCTP, Pohang for warm hospitality. CP was supported by Basic Science Research Program through the National Research Foundation of Korea funded by the Ministry of Education (NRF-2016R1D1A1B03932371). SS is supported by Basic Science Research Program through the National Research Foundation of Korea (NRF-2017R1D1A1B03034222).
%
%
\appendix
\def\thesection{Appendix \Alph{section}}
\setcounter{equation}{0}
\renewcommand{\theequation}{\Alph{section}.\arabic{equation}}
\section{Vacuum structures}\label{sec:app_vac_str}
In this appendix, we prove (\ref{eq:4m-3}), (\ref{eq:4m-2}), (\ref{eq:4m-1}) and (\ref{eq:4m}). The vacuum structures that are connected to the maximum number of elementary walls in the nonlinear sigma models on $SO(2N)/U(N)$ are studied by decomposing the diagrams into two-dimensional diagrams in \cite{Lee:2017kaj}. We use the same method in the nonlinear sigma models on $Sp(N)/U(N)$. The rule for the decomposition is that the simple roots that have already appeared in the previous diagrams should not be repeated.

\begin{figure}[ht!]
\begin{center}
$\begin{array}{ccc}
\includegraphics[width=6cm,clip]{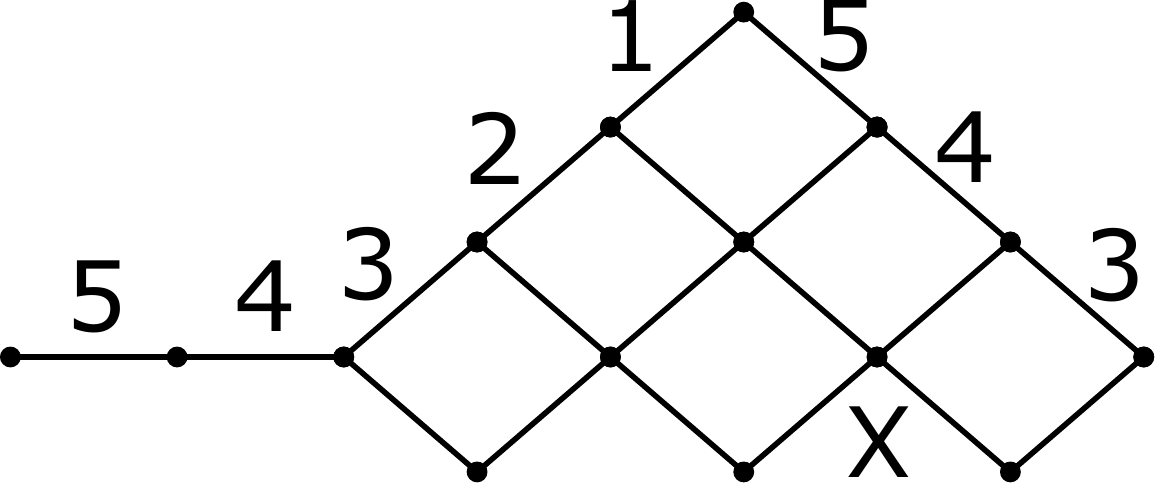}
&~
&
\includegraphics[width=2cm,clip]{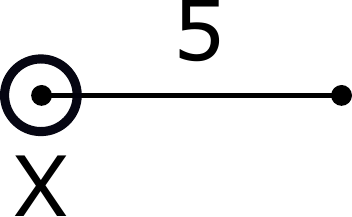}\\
\mathrm{(a)} &~~~~~~~& \mathrm{(b)}
\end{array}
$
\end{center}
 \caption{$N=5$ case. The vacuum structure near $\la 1 \ra$ ($\la 32\ra$) decomposes into two diagrams. }
 \label{fig:n5decomp}
\end{figure}

The vacuum structure of $N=5$ case is depicted in Figure \ref{fig:n5}. The vacuum structure near $\la 1 \ra$($\la 32 \ra$) decomposes into two diagrams, as is shown in Figure \ref{fig:n5decomp}. The circle indicates $\la A \ra$($\la B \ra$). The letter `X' indicates the vacuum that is connected to the both diagrams. The left-hand side(the right-hand side) of the each diagram is the limit as $x\to +\infty(x\to-\infty)$ for the vacuum structure near $\la 1 \ra$ whereas the left-hand side(the right-hand side) of the each diagrams is the limit as $x\to -\infty(x\to+\infty)$ for the vacuum structure near $\la 32 \ra$.

\begin{figure}[ht!]
\begin{center}
$\begin{array}{ccc}
\includegraphics[width=6cm,clip]{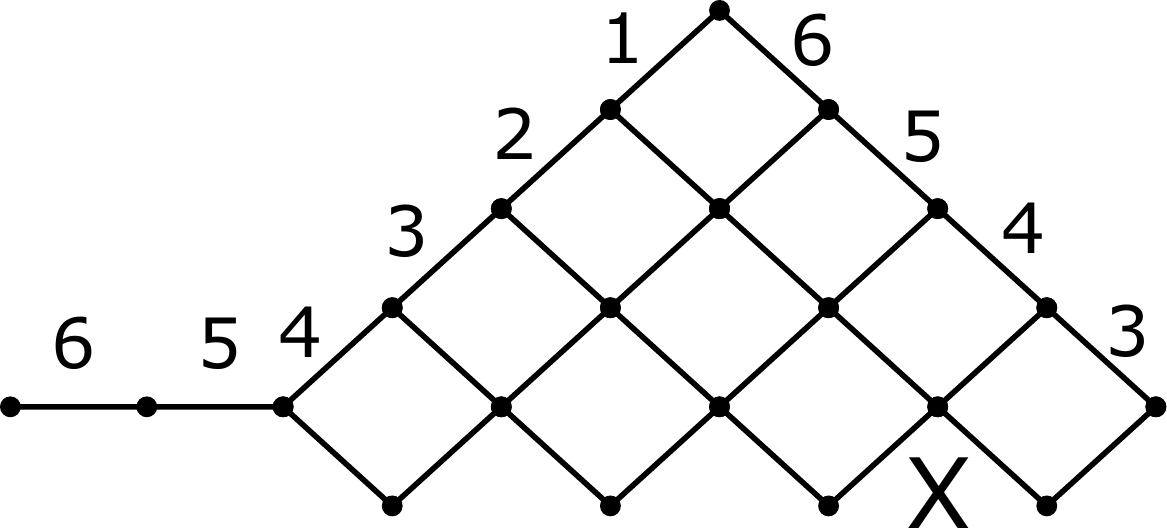}
&~
&
\includegraphics[width=5cm,clip]{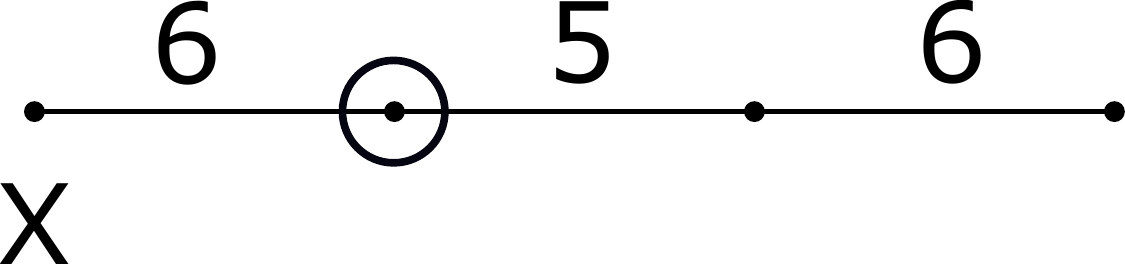}\\
\mathrm{(a)} &~~~~~~~& \mathrm{(b)}
\end{array}
$
\end{center}
 \caption{$N=6$ case. The vacuum structure near $\la 1 \ra$ ($\la 64\ra$) decomposes into two diagrams. }
 \label{fig:n6decomp}
\end{figure}

\begin{figure}[ht!]
\begin{center}
$\begin{array}{ccc}
\includegraphics[width=6cm,clip]{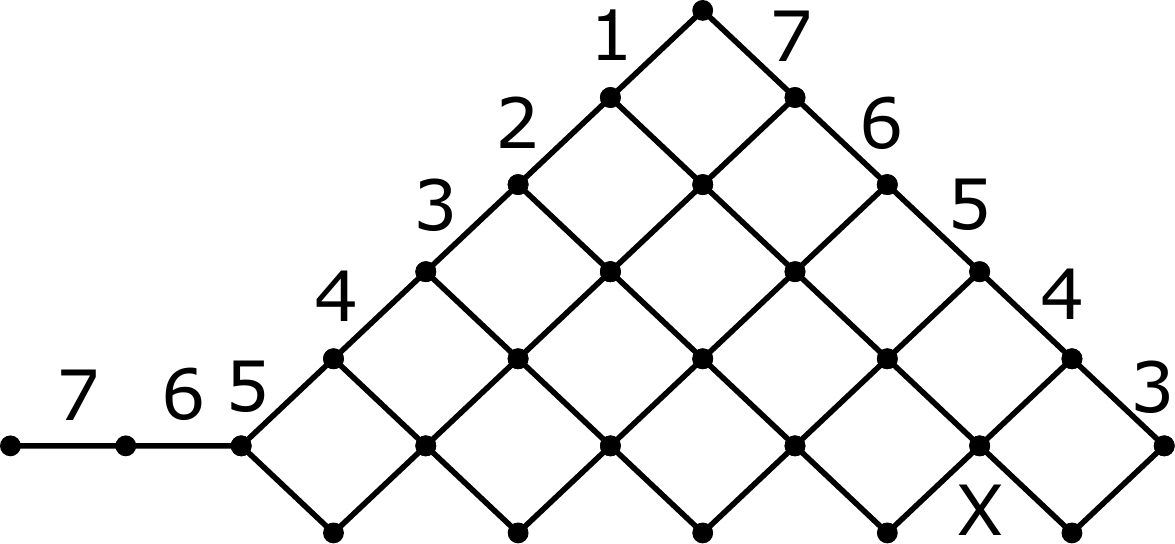}
&~
&
\includegraphics[width=6cm,clip]{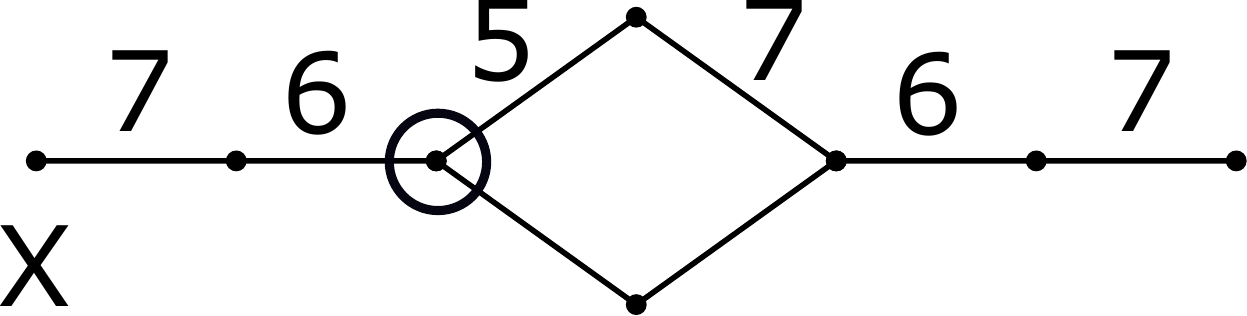}\\
\mathrm{(a)} &~~~~~~~& \mathrm{(b)}
\end{array}
$
\end{center}
 \caption{$N=7$ case. The vacuum structure near $\la 1 \ra$ ($\la 128\ra$) decomposes into two diagrams. }
  \label{fig:n7decomp}
\end{figure}

\begin{figure}[ht!]
\begin{center}
$\begin{array}{ccc}
\includegraphics[width=6cm,clip]{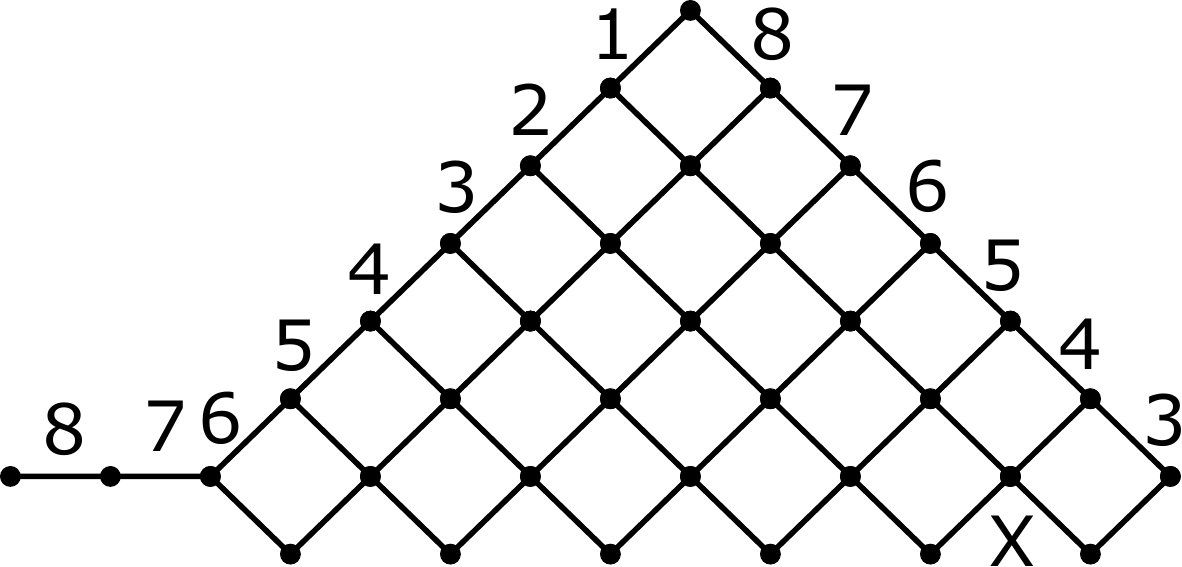}
&~
&
\includegraphics[width=6cm,clip]{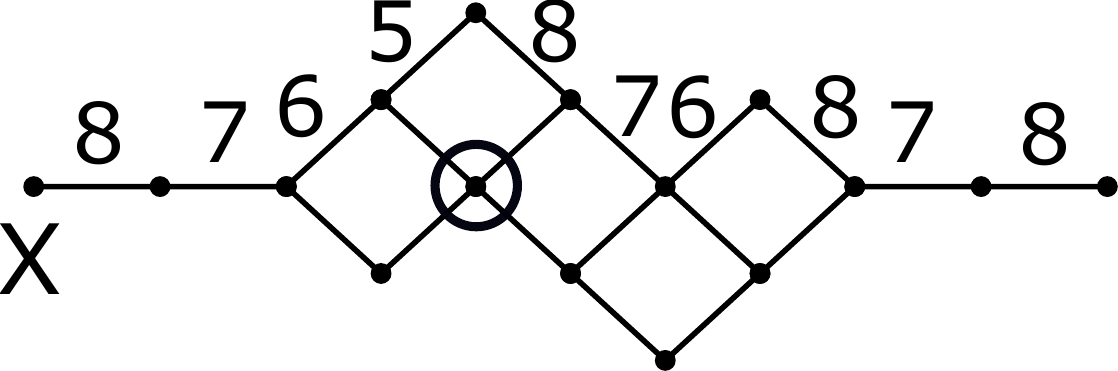}\\
\mathrm{(a)} &~~~~~~~& \mathrm{(b)}
\end{array}
$
\end{center}
 \caption{$N=8$ case. The vacuum structure near $\la 1 \ra$ ($\la 256\ra$) decomposes into two diagrams. }
  \label{fig:n8decomp}
\end{figure}

Figure \ref{fig:n6}, which describes the vacuum structure of $N=6$ case decomposes into two diagrams as is shown in Figure \ref{fig:n6decomp}. In the same manner the vacuum structures of $N=7$ and $N=8$ cases are presented in Figure \ref{fig:n7decomp} and Figure \ref{fig:n8decomp}. The vacuum structures repeat the four diagrams in Figure \ref{fig:n1n2n3n4}.

\begin{figure}[ht!]
\begin{center}
$\begin{array}{ccc}
\includegraphics[width=6cm,clip]{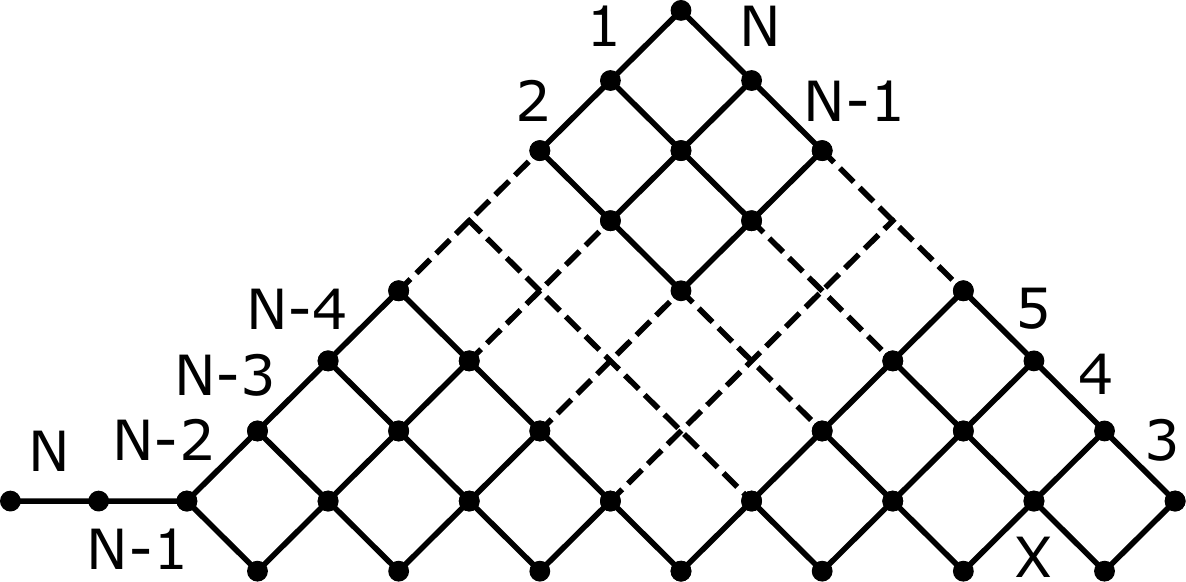}
&~
&
\includegraphics[width=6cm,clip]{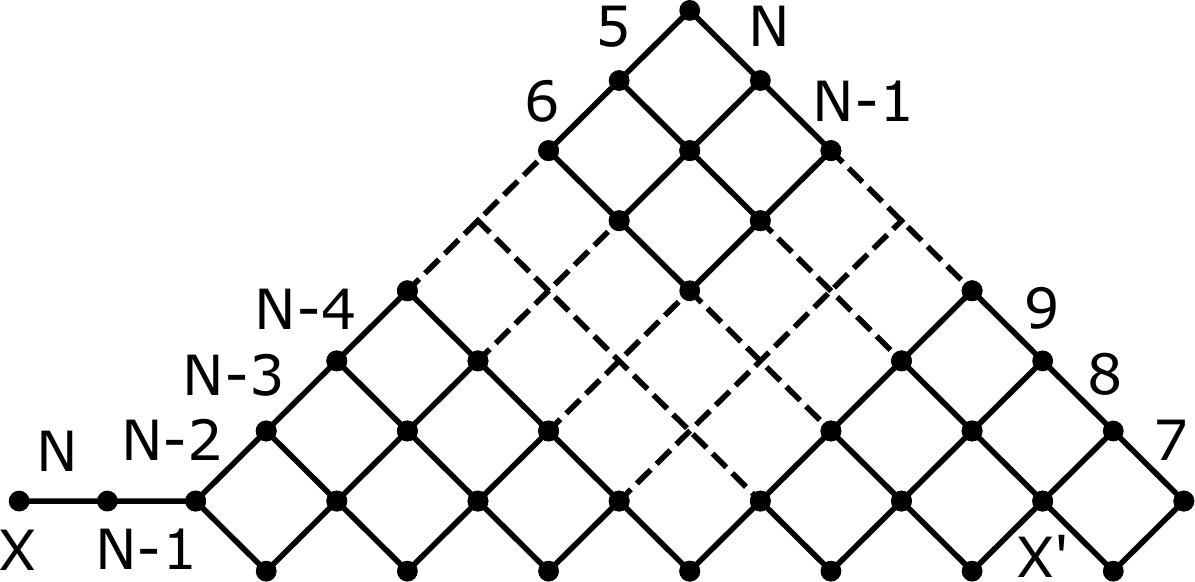}\\
\mathrm{(a)} &~~~~~~~& \mathrm{(b)}
\end{array}
$
\end{center}
 \caption{First two diagrams of the vacuum structure near $\la 1 \ra$ and $\la 2^N \ra$.  }
 \label{fig:1st2}
\end{figure}

\begin{figure}[ht!]
\begin{center}
$\begin{array}{ccc}
\includegraphics[width=3cm,clip]{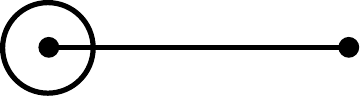}
&~~~~~~&
\includegraphics[width=6cm,clip]{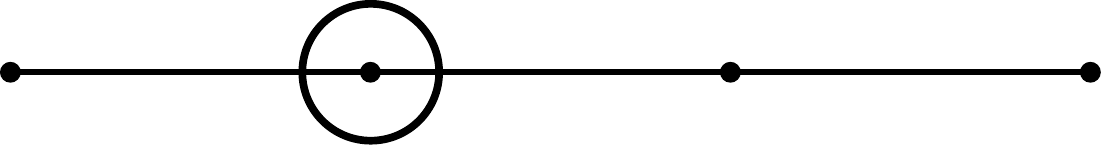}\vspace{1cm}\\
\mathrm{(a)} &~~~~~~& \mathrm{(b)}\\
\includegraphics[width=6cm,clip]{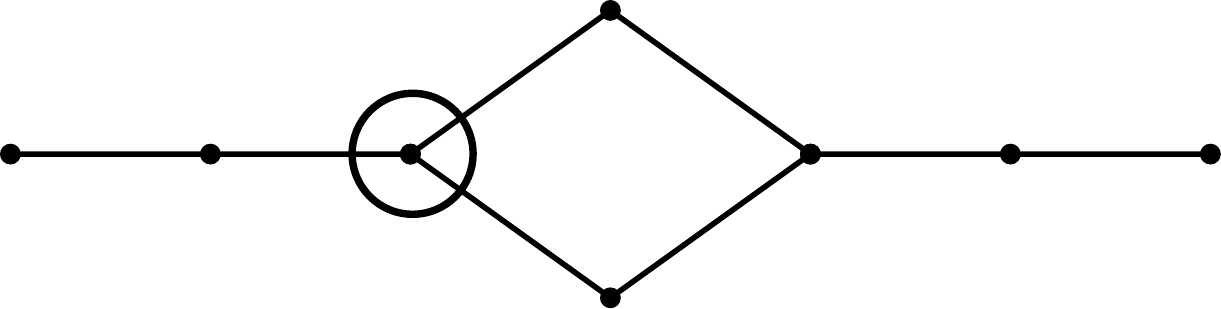}
&~~~~~~&
\includegraphics[width=7cm,clip]{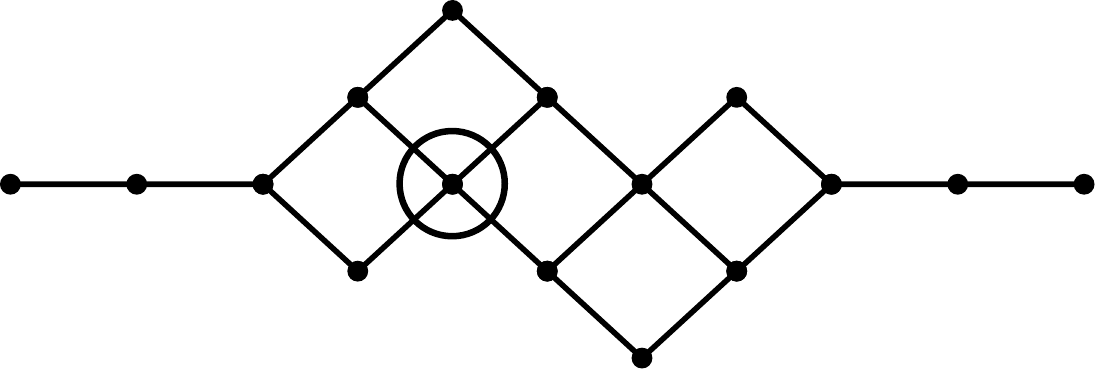}\vspace{1cm}\\
\mathrm{(c)} &~~~~~~& \mathrm{(d)}
\end{array}
$
\end{center}
 \caption{Four types of vacuum structures. The circle indicates the vacuum that is connected to the maximum number of simple roots.  }
 \label{fig:4diagrams}
\end{figure}

All the vacuum structures can be decomposed into two dimensional diagrams in Figure \ref{fig:1st2}, where only the first two diagrams are shown and then fall into four categories. The vacuum that is connected to the maximum number of elementary walls is circled in each diagram in Figure \ref{fig:4diagrams}.

\begin{figure}[ht!]
\begin{center}
\includegraphics[width=12cm,clip]{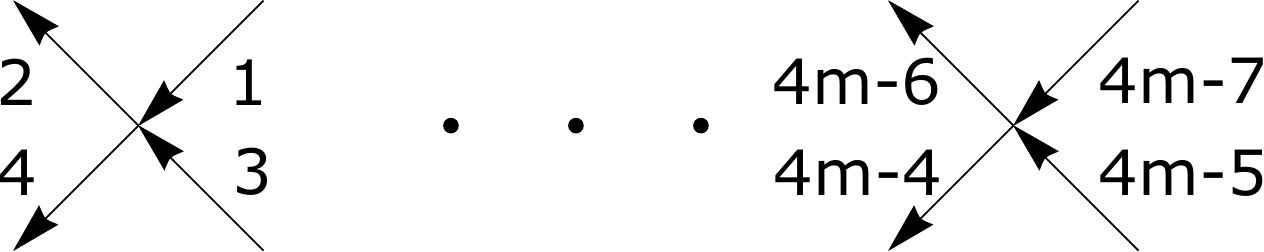}
\end{center}
 \caption{common part near $\la A \ra$ }
 \label{fig:commonpt_a}
\end{figure}

\begin{figure}[ht!]
\begin{center}
\includegraphics[width=12cm,clip]{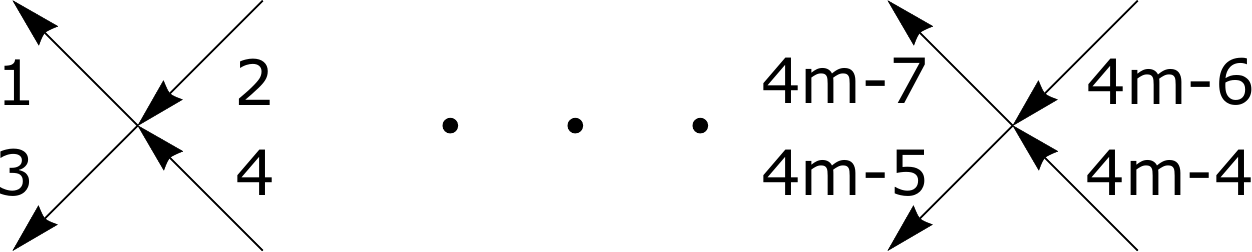}
\end{center}
 \caption{common part near $\la B \ra$ }
 \label{fig:commonpt_b}
\end{figure}

\begin{figure}[ht!]
\begin{center}
$\begin{array}{ccc}
\includegraphics[width=6cm,clip]{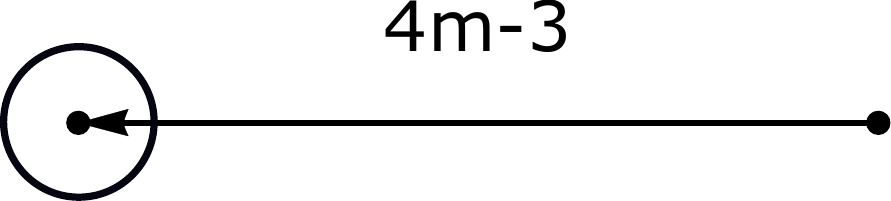}
&~~~~~~&
\includegraphics[width=6cm,clip]{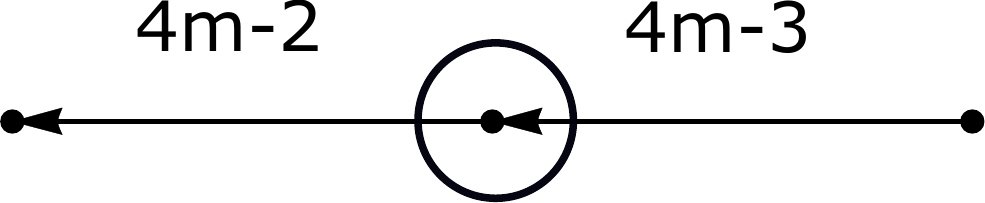}\\
~~ & ~~ & ~~ \\
\mathrm{(a)} &~~~~~~& \mathrm{(b)}  \\
\includegraphics[width=6cm,clip]{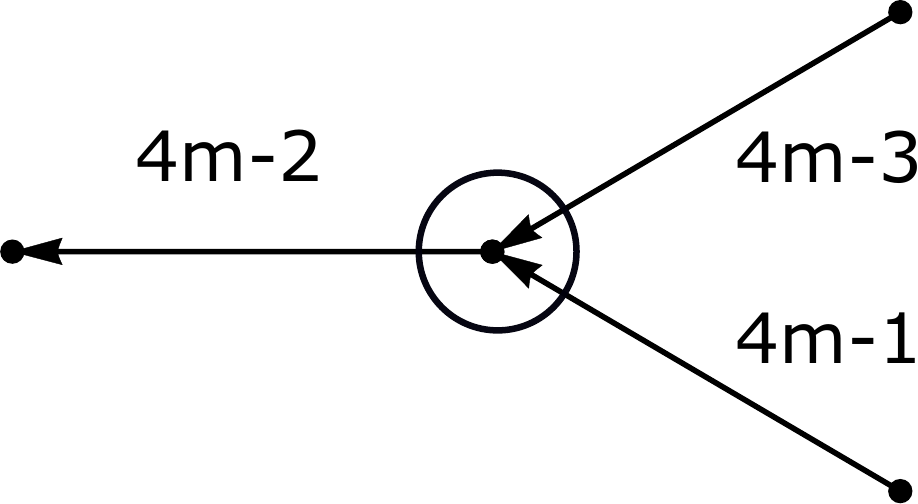}
&~~~~~~&
\includegraphics[width=6cm,clip]{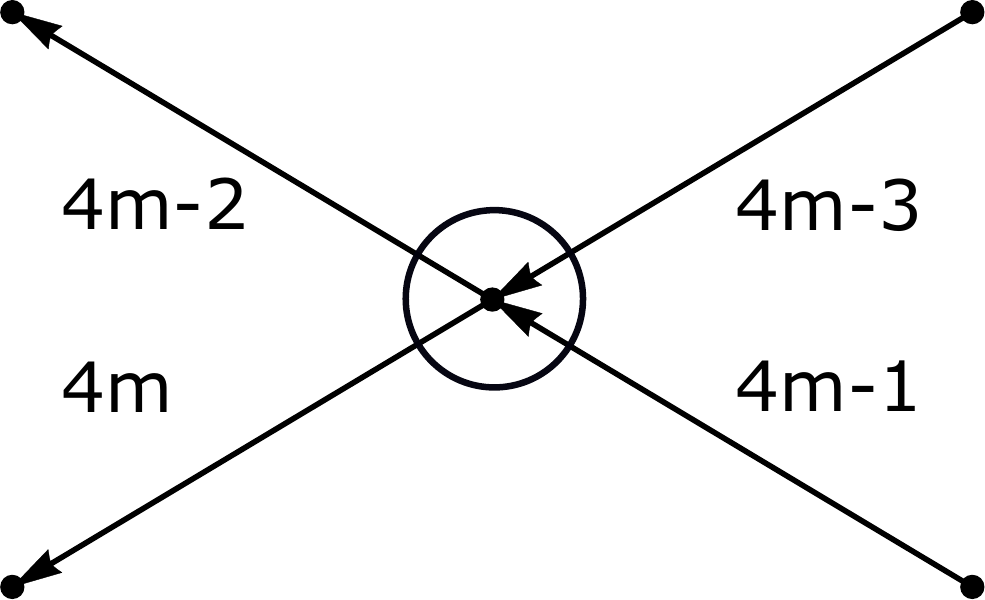}\\
~~ & ~~ & ~~ \\
\mathrm{(c)} &~~~~~~& \mathrm{(d)}
\end{array}
$
\end{center}
 \caption{Remaining part of the vacuum structure near $\la A \ra$. The location of $\la A \ra$ is circled.}
 \label{fig:restpt_a}
\end{figure}

\begin{figure}[ht!]
\begin{center}
$\begin{array}{ccc}
\includegraphics[width=6cm,clip]{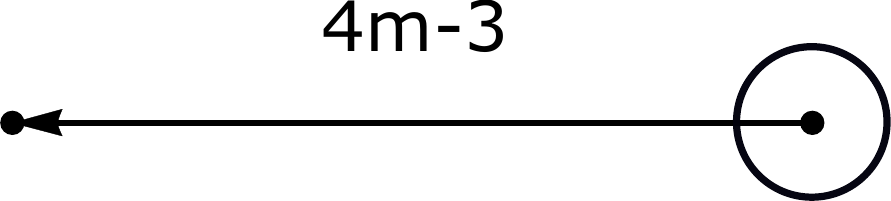}
&~~~~~~&
\includegraphics[width=6cm,clip]{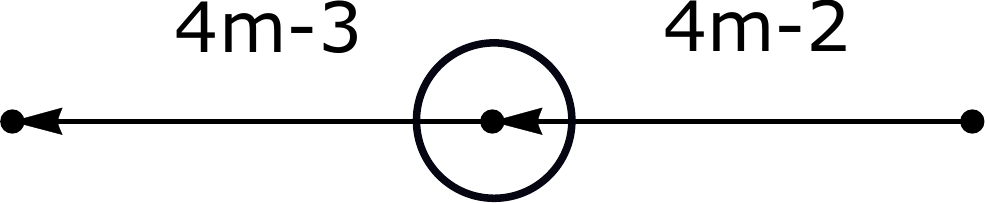}\\
~~ & ~~ & ~~ \\
\mathrm{(a)} &~~~~~~& \mathrm{(b)}  \\
\includegraphics[width=6cm,clip]{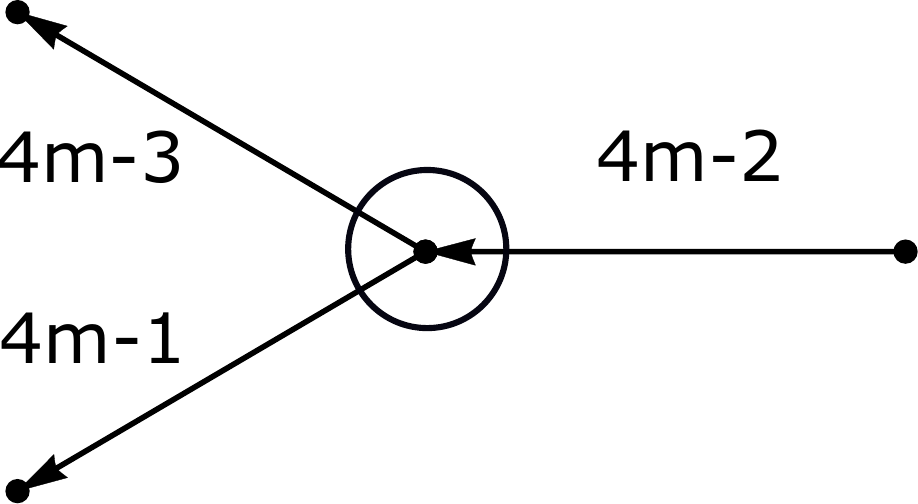}
&~~~~~~&
\includegraphics[width=6cm,clip]{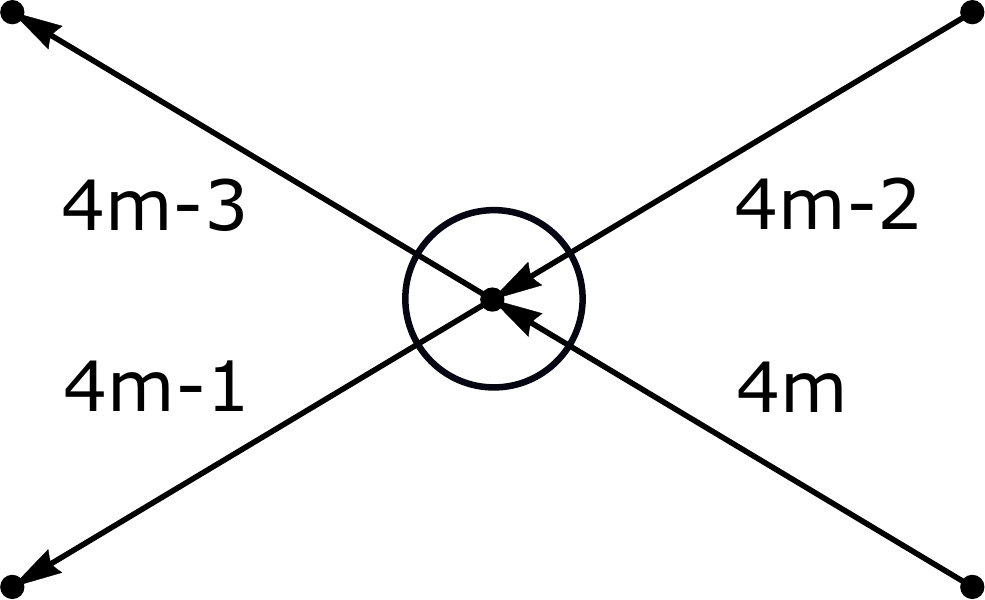}\\
~~ & ~~ & ~~ \\
\mathrm{(c)} &~~~~~~& \mathrm{(d)}
\end{array}
$
\end{center}
 \caption{Remaining part of the vacuum structure near $\la B \ra$. The location of $\la B \ra$ is circled.}
 \label{fig:restpt_b}
\end{figure}

The vacuum structures that are connected to the maximum number of elementary walls can be obtained from the repeated diagrams. $\la A \ra$ denotes the vacuum near $\la 1 \ra$ and $\la B \ra$ denotes the vacuum near $\la 2^N \ra$. The common parts of each vacuum structure near $\la A \ra$ and $\la B \ra$ are shown in Figure \ref{fig:commonpt_a} and Figure \ref{fig:commonpt_b}. The rest of the vacuum structures are obtained from Figure \ref{fig:4diagrams}. The remaining parts of each vacuum structure near $\la A \ra$ and $\la B \ra$  are shown in Figure \ref{fig:restpt_a} and Figure \ref{fig:restpt_b} for $N=4m-3$, $N=4m-2$, $N=4m-1$ and $N=4m$.

The vacuum structure of $\la A \ra$ is derived from Figure \ref{fig:commonpt_a} and Figure \ref{fig:restpt_a} as follows:

\bul $N=4m-3, ~(m\geq2)$
\bea
\Big\{\underbrace{\vec{2},\vec{4},\cdots,\overrightarrow{4m-6},\overrightarrow{4m-4}}_{2m-2}\Big\} \leftarrow \la A \ra \leftarrow \Big\{\underbrace{\vec{1},\vec{3},\cdots,\overrightarrow{4m-5},\overrightarrow{4m-3}}_{2m-1}\Big\}
\label{eq:4m-3a}
\eea

\bul $N=4m-2, ~(m\geq2)$
\bea
\Big\{\underbrace{\vec{2},\vec{4},\cdots,\overrightarrow{4m-4},\overrightarrow{4m-2}}_{2m-1}\Big\} \leftarrow \la A \ra \leftarrow \Big\{\underbrace{\vec{1},\vec{3},\cdots,\overrightarrow{4m-5},\overrightarrow{4m-3}}_{2m-1}\Big\} \label{eq:4m-2a}
\eea

\bul $N=4m-1, ~(m\geq2)$
\bea
\Big\{\underbrace{\vec{2},\vec{4},\cdots,\overrightarrow{4m-4},\overrightarrow{4m-2}}_{2m-1}\Big\} \leftarrow \la A \ra \leftarrow \Big\{\underbrace{\vec{1},\vec{3},\cdots,\overrightarrow{4m-3},\overrightarrow{4m-1}}_{2m}\Big\} \label{eq:4m-1a}
\eea

\bul $N=4m, ~(m\geq2)$
\bea
\Big\{\underbrace{\vec{2},\vec{4},\cdots,\overrightarrow{4m-2},\overrightarrow{4m}}_{2m}\Big\} \leftarrow \la A \ra \leftarrow \Big\{\underbrace{\vec{1},\vec{3},\cdots,\overrightarrow{4m-3},\overrightarrow{4m-1}}_{2m}\Big\} \label{eq:4ma}
\eea
Each case with $m=1$ is shown in Figure \ref{fig:n1n2n3n4}. Each case with $m=2$ is shown in Figure \ref{fig:n5decomp}, Figure \ref{fig:n6decomp}, Figure \ref{fig:n7decomp} and Figure \ref{fig:n8decomp}. Let us assume that (\ref{eq:4m-3a}), (\ref{eq:4m-2a}),
(\ref{eq:4m-1a}) and (\ref{eq:4ma}) are true. Then these are true for $m^\prime=m+1$ as it corresponds to adding one more diagram in Figure \ref{fig:commonpt_a}. Therefore (\ref{eq:4m-3a}), (\ref{eq:4m-2a}),
(\ref{eq:4m-1a}) and (\ref{eq:4ma}) are true.

The vacuum structure of $\la B \ra$ is derived from Figure \ref{fig:commonpt_b} and Figure \ref{fig:restpt_b} as follows:

\bul $N=4m-3, ~(m\geq2)$
\bea
\Big\{\underbrace{\vec{1},\vec{3},\cdots,\overrightarrow{4m-5},\overrightarrow{4m-3}}_{2m-1}\Big\}
\leftarrow \la B \ra \leftarrow \Big\{\underbrace{\vec{2},\vec{4},\cdots,\overrightarrow{4m-6},\overrightarrow{4m-4}}_{2m-2}\Big\} \label{eq:4m-3b}
\eea

\bul $N=4m-2, ~(m\geq2)$
\bea
\Big\{\underbrace{\vec{1},\vec{3},\cdots,\overrightarrow{4m-5},\overrightarrow{4m-3}}_{2m-1}\Big\}
\leftarrow \la B \ra \leftarrow
\Big\{\underbrace{\vec{2},\vec{4},\cdots,\overrightarrow{4m-4},\overrightarrow{4m-2}}_{2m-1}\Big\}
\label{eq:4m-2b}
\eea

\bul $N=4m-1, ~(m\geq2)$
\bea
\Big\{\underbrace{\vec{1},\vec{3},\cdots,\overrightarrow{4m-3},\overrightarrow{4m-1}}_{2m}\Big\}\leftarrow \la B \ra \leftarrow
\Big\{\underbrace{\vec{2},\vec{4},\cdots,\overrightarrow{4m-4},\overrightarrow{4m-2}}_{2m-1}\Big\}
\label{eq:4m-1b}
\eea

\bul $N=4m, ~(m\geq2)$
\bea
\Big\{\underbrace{\vec{1},\vec{3},\cdots,\overrightarrow{4m-3},\overrightarrow{4m-1}}_{2m}\Big\}
\leftarrow \la B \ra \leftarrow
\Big\{\underbrace{\vec{2},\vec{4},\cdots,\overrightarrow{4m-2},\overrightarrow{4m}}_{2m}\Big\}
\label{eq:4mb}
\eea
Each case with $m=1$ is shown in Figure \ref{fig:n1n2n3n4}. Each case with $m=2$ is shown in Figure \ref{fig:n5decomp}, Figure \ref{fig:n6decomp}, Figure \ref{fig:n7decomp} and Figure \ref{fig:n8decomp}. Let us assume that (\ref{eq:4m-3b}), (\ref{eq:4m-2b}),
(\ref{eq:4m-1b}) and (\ref{eq:4mb}) are true. Then these are true for $m^\prime=m+1$ as it corresponds to adding one more diagram in Figure \ref{fig:commonpt_b}. Therefore (\ref{eq:4m-3b}), (\ref{eq:4m-2b}),
(\ref{eq:4m-1b}) and (\ref{eq:4mb}) are true.

For any $N$ the vacuum structure is
\bea
\vec{N} \leftarrow \cdots\leftarrow \la A \ra \leftarrow \cdots \leftarrow
\la B \ra \leftarrow \cdots \leftarrow \vec{N}
\eea
Therefore (\ref{eq:4m-3}), (\ref{eq:4m-2}), (\ref{eq:4m-1}) and (\ref{eq:4m}) are proved.

\end{document}